\documentclass[12pt]{article}
\usepackage{amsfonts, amsthm, amsmath}
\usepackage{bm}
\usepackage[utf8]{inputenc}
\usepackage[top=2cm, left=2cm,right=2cm, bottom=3cm]{geometry}
\usepackage{amssymb}
\usepackage{graphicx}
\usepackage{xcolor}
\usepackage{cite}
\numberwithin{equation}{section}

\usepackage[colorlinks=true,
            linkcolor=blue,
            citecolor=blue,
            urlcolor=blue]{hyperref}

\allowdisplaybreaks

\numberwithin{equation}{section}

\newtheorem{theorem}{Theorem}
\newtheorem{proposition}[theorem]{Proposition}

\allowdisplaybreaks

\title{ {\bf
Superintegrable systems of two interacting spin-$\frac12$ particles with first-order vector integrals of motion
}}

\author{
        {\bf O. O\u{g}ulcan Tuncer}\thanks{E-mail address:
        otuncer@hacettepe.edu.tr}\:\:\footnote{Corresponding Author}\,, \,
        { \bf \.{I}smet Yurdu\c{s}en}\thanks{E-mail address:
       yurdusen@hacettepe.edu.tr}
 \\
  \\Department of Mathematics, Hacettepe University,
                    \\ 06800 Beytepe, Ankara, Turkey}

\date{}
\begin{document}

\maketitle

\begin{abstract}
We study quantum superintegrability for two interacting non-relativistic spin-$\frac12$ particles in three-dimensional Euclidean space. The Hamiltonian contains a central potential together with spin-orbit, spin-spin, tensor, and quadratic spin-orbit interaction terms, all depending only on the relative distance. We restrict the classification to the $V_4=0$ case, so that the spin-momentum interaction term is not included. We determine all such systems admitting non-trivial first-order vector integrals of motion. For this purpose, we construct the most general Hermitian first-order vector operator built from the relative position, momentum, orbital angular momentum, and the two spin vectors. The commutativity condition with the Hamiltonian leads to an overdetermined system of radial determining equations, whose solution gives the complete list of admissible potentials and the corresponding vector integrals within this class. The results extend the previous classifications of scalar and pseudo-scalar first-order integrals for two particles with spin. We also discuss selected symmetry algebras generated by the vector integrals and show,
in one representative case, how a scalar reduction leads to exact Coulomb- and oscillator-type solutions.\medskip\\
\textit{Keywords:} superintegrable systems, spin, vector integrals of motion, nucleon-nucleon interaction, symmetry algebra, exactly solvable models\medskip\\
\textit{PACS numbers:} 02.30.Ik, 03.65.-w, 11.30.-j, 13.75.Cs
\end{abstract}

\section{Introduction}

The systematic study of integrable and superintegrable Hamiltonian systems has
played an important role in mathematical physics for several decades. In
classical mechanics, a Hamiltonian system with $n$ degrees of freedom is
called integrable if it admits $n$ functionally independent integrals of motion
in involution, including the Hamiltonian. It is called superintegrable if it
possesses further functionally independent integrals of motion. In the quantum
case, the corresponding definition is formulated in terms of algebraically
independent operators commuting with the Hamiltonian. Superintegrable systems
are of particular interest because their additional symmetries often lead to
exact solvability, degeneracies in the spectrum, and nontrivial symmetry
algebras.

For natural Hamiltonians of the form
\[
H=-\frac{\hbar^2}{2}\Delta+V_0(\vec{x}),
\]
the classification of classical and quantum superintegrable systems with
integrals polynomial in the momenta goes back to the foundational works
\cite{Fris,Makarov}. Further systematic studies led to the classification of
many important families of superintegrable systems in Euclidean spaces and in
spaces of constant or nonconstant curvature
\cite{Evans.a,Evans.b,Grosche2,Kalnins.d,Kalnins.f,Rodriguez,Kalnins.h,Kalnins.i}.
The Kepler--Coulomb system and the isotropic harmonic oscillator are the most
prominent spherically symmetric examples; they are also singled out by
Bertrand's theorem as the only central potentials for which all bounded
classical trajectories are closed \cite{Bertrand}. Higher-order
superintegrability and the associated polynomial algebras have also been
studied extensively
\cite{Gravel.a,Tremblay.b,Marquette.b,Tremblay.c,PostWinternitz:2015,MSW,EWY}.
For a broad overview of classical and quantum superintegrability and its
applications, see \cite{MillerPostWinternitz:2013}.

Another important direction is the study of Hamiltonian systems with
velocity-dependent or momentum-dependent interaction terms. This includes
integrable and superintegrable systems in magnetic fields
\cite{Dorizzi,Berube,Libor2,Libor4,Libor5,Libor6,EM2026}. Spin-dependent
Hamiltonians provide a different but closely related extension of the spinless
theory. In such systems, the spin degrees of freedom enlarge the set of
admissible interaction terms and give rise to scalar, pseudo-scalar, vector,
axial-vector, tensor, and pseudo-tensor integrals of motion.

The systematic study of superintegrability with spin was initiated for systems
involving one spin-$1/2$ particle interacting with a spinless particle
\cite{Winternitz.c}. The three-dimensional case with first-order integrals was
studied in \cite{wy3}, and higher-order tensorial structures were later
investigated in \cite{DWY,YTW}. Related approaches to superintegrable systems
with arbitrary spin and matrix potentials can be found in
\cite{Pronko.b,Nikitin.e,Nikitin.f}. Other physically motivated examples include
spin-$1/2$ particles interacting with dyons and monopoles
\cite{DHoker,Feher}.

A natural next step is to consider the interaction of two non-relativistic
particles, both with spin $1/2$. This setting is particularly rich because it
contains the standard spin-spin, spin-orbit, tensor, spin-momentum, and
quadratic spin-orbit terms appearing in the Okubo--Marshak form of the
two-nucleon interaction \cite{OM}. In recent work, this program was initiated by
classifying all such two-spin systems admitting additional first-order scalar
integrals of motion \cite{TuncerYurdusen2025}. It was subsequently continued by
the classification of first-order pseudo-scalar integrals
\cite{TurkkanTuncerYurdusen2026}.

The present paper continues this research program by considering first-order
vector integrals of motion. This is a natural and important step for two
reasons. First, vector integrals form a substantially larger and more structured
class of symmetries than scalar or pseudo-scalar ones. Secondly, vector
integrals often encode hidden symmetry structures: their commutation relations
with the angular momentum and with one another may generate nontrivial symmetry
algebras, and suitable scalar combinations of vector integrals can be used as
additional commuting observables. Thus vector integrals are not only useful for
classification, but also for understanding the algebraic and spectral structure
of the corresponding quantum systems.

We consider the most general rotationally invariant Hamiltonian for two
interacting spin-$1/2$ particles with spherically symmetric radial
coefficients. In the full off-shell Okubo--Marshak potential, the spin-momentum
term
\[
V_4(r)(\vec{\sigma}_1,\vec{p})(\vec{\sigma}_2,\vec{p})
\]
is an independent interaction. On the energy shell, however, this term can be
eliminated in favour of the remaining spin-dependent structures. Motivated by
this reduction, and in order to keep the determining equations tractable, we
restrict the present classification to the subfamily $V_4(r)=0$. The remaining
Hamiltonian still contains central, spin-orbit, spin-spin, tensor, and
quadratic spin-orbit interactions, and is sufficiently general to exhibit a wide
variety of vector-superintegrable structures.

Our goal is to determine all choices of the radial potentials for which this
Hamiltonian admits an additional non-trivial first-order vector integral of
motion. We construct the most general Hermitian first-order vector operator from
the basic vectorial directions generated by
$\vec{x}$, $\vec{p}$, $\vec{L}$, $\vec{\sigma}_1$, and
$\vec{\sigma}_2$. The symmetrization of the resulting operator products is
performed using the procedure developed in \cite{YT}. Imposing the commutativity
condition
\[
[H,\vec X]=0
\]
and equating the coefficients of the independent differential operators gives a large overdetermined system of radial determining equations for the coefficient functions of $\vec X$ and for the potentials. Solving this system gives the complete classification, within the $V_4=0$ class, stated in Theorem~1.

The full step-by-step elimination of the determining equations is not repeated
here. The computation is carried out with Mathematica, following the standard
procedure used in the scalar and pseudo-scalar classifications
\cite{TuncerYurdusen2025,TurkkanTuncerYurdusen2026}. In those works, the
method was described in detail: one substitutes the general ansatz into the
commutation condition, separates the resulting expression with respect to
independent differential operators, and solves the resulting overdetermined
system by successive case distinctions.

For completeness, the determining equations obtained from this procedure are listed in Appendix~A. We do not include the full branch-by-branch elimination, since it is long and follows the same pattern as in the previous classifications. Instead, in the main text we give the general ansatz and state the complete list of admissible potentials together with the corresponding vector integrals. Each family can be checked directly by substitution into $[H,\vec X]=0.$

Beyond the classification itself, we also discuss the algebraic and spectral
roles of selected vector integrals. In particular, some of the vector integrals
generate closed symmetry algebras together with the total angular momentum or
lead to natural scalar operators by contraction with $\vec J$. For one
representative family, this scalar reduction gives a complete separation of the
spin-angular variables and leads to exact solutions for Coulomb and harmonic
oscillator choices of the central potential.

The paper is organized as follows. In Section~2 we formulate the two-spin
Hamiltonian and specify the $V_4=0$ subfamily considered in this work. In
Section~3 we construct the most general first-order vector integral and state
the complete classification theorem. Section~4 is devoted to symmetry algebras
associated with selected cases. In Section~5 we present an exact solution
obtained from a scalar reduction of a first-order vector integral. Finally,
Section~6 contains concluding remarks and possible directions for future work. The determining equations are collected in Appendix~A.

\section{Formulation of the problem}

We consider two interacting non-relativistic particles, both with spin $1/2$,
moving in the three-dimensional Euclidean space $E^3$. Let
$\vec{x}^{(i)}$, $\vec{p}^{(i)}$, and $\vec{\sigma}_i$, $i=1,2$, denote
the position, momentum, and Pauli spin operators of the two particles,
respectively. Passing to relative variables, we set
\[
\vec{x}=\vec{x}^{(1)}-\vec{x}^{(2)},\qquad
\vec{p}=\frac12\bigl(\vec{p}^{(1)}-\vec{p}^{(2)}\bigr),
\qquad
r=\|\vec{x}\|.
\]

The most general interaction potential for two spin-$1/2$ particles is
obtained by imposing translational invariance, Galilean invariance, permutation
symmetry, rotational invariance, space-reflection invariance, time-reversal
invariance, and Hermiticity. Under these assumptions, the interaction is of
Okubo--Marshak type and is built from the relative variables
$\vec{x}$, $\vec{p}$, $\vec{\sigma}_1$, and $\vec{\sigma}_2$.

Restricting to spherically symmetric coefficient functions, the Hamiltonian is
of the form
\begin{equation}\label{eq:general_H_full}
H=-\frac{\hbar^2}{2}\Delta+V,
\end{equation}
where
\begin{align}
V={}&V_0(r)
+\frac12 V_1(r)\bigl(\vec{\sigma}_1+\vec{\sigma}_2,\vec{L}\bigr)
+V_2(r)(\vec{\sigma}_1,\vec{\sigma}_2)
+V_3(r)(\vec{x},\vec{\sigma}_1)(\vec{x},\vec{\sigma}_2)
\nonumber\\
&+V_4(r)(\vec{\sigma}_1,\vec{p})(\vec{\sigma}_2,\vec{p})
+\frac12 V_5(r)\Bigl(
(\vec{\sigma}_1,\vec{L})(\vec{\sigma}_2,\vec{L})
+
(\vec{\sigma}_2,\vec{L})(\vec{\sigma}_1,\vec{L})
\Bigr).
\label{eq:general_V_full}
\end{align}
Here
\[
\vec{L}=\vec{x}\wedge\vec{p}
\]
is the orbital angular momentum operator. We use the convention
\[
p_k=-i\hbar\,\partial_{x_k},\qquad
\vec{\sigma}_1=\vec{\sigma}\otimes I_2,\qquad
\vec{\sigma}_2=I_2\otimes\vec{\sigma},
\]
so that the Hamiltonian acts on four-component spinors.

The term
\[
V_4(r)(\vec{\sigma}_1,\vec{p})(\vec{\sigma}_2,\vec{p})
\]
is the spin-momentum interaction. In the off-shell formulation it is an
independent part of the Okubo--Marshak potential. On the energy shell, however,
this term can be eliminated in favour of the remaining spin-dependent
structures. Motivated by this reduction, and in order to keep the determining
system tractable, in the present paper we restrict the classification to the
subfamily
\[
V_4(r)=0.
\]

The total angular momentum is
\[
\vec{J}=\vec{L}+\hbar\vec{S},
\qquad
\vec{S}=\frac12(\vec{\sigma}_1+\vec{\sigma}_2).
\]
Since the Hamiltonian is rotationally invariant, we have
\[
[H,J^2]=0,\qquad [H,J_3]=0.
\]
Consequently,
\[
H,\qquad J^2,\qquad J_3
\]
form a commuting set of integrals of motion, and the system is integrable by
construction.

We also note that the spin-exchange scalar
\[
K=(\vec{\sigma}_1,\vec{\sigma}_2)
\]
commutes with the Hamiltonian for arbitrary choices of the radial potentials.
This operator only distinguishes the singlet and triplet spin sectors and will
therefore be regarded as a trivial integral of motion.

Finally, one has to take into account the potentials which are induced from
scalar Hamiltonians by unitary gauge transformations. Such potentials do not
represent genuinely new spin-dependent interactions, since they are obtained
from spin-independent systems by a change of gauge.

For the two-spin Hamiltonian considered here, the gauge-induced potentials are
\begin{equation}
V_1(r)=\frac{2\hbar}{r^2},\qquad
V_2(r)=\frac{\hbar^2}{r^2},\qquad
V_3(r)=-\frac{\hbar^2}{r^4},\qquad
V_4(r)=V_5(r)=0,
\label{eq:gauge_induced_potentials}
\end{equation}
with $V_0(r)$ arbitrary. The corresponding integrals of motion are obtained by applying the same
unitary gauge transformation to the integrals of the underlying scalar
Hamiltonian, see \cite{TuncerYurdusen2025} for details. Therefore, whenever the potentials in
\eqref{eq:gauge_induced_potentials} occur in the classification below, the
system will be identified as gauge induced and separated from the genuinely new
vector-superintegrable families.

The aim of the present paper is to determine all spherically symmetric
potentials $V_i(r)$, with $V_4(r)=0$, for which the Hamiltonian
\eqref{eq:general_H_full} admits additional non-trivial first-order vector integrals
of motion.

\section{First-order vector integrals of motion}

In this section we determine the spherically symmetric Hamiltonians of the form
\eqref{eq:general_H_full} with $V_4=0$ which admit additional first-order vector integrals
of motion. By a vector integral we mean a vector operator
\[
\vec X=(X_1,X_2,X_3)
\]
whose components commute with the Hamiltonian,
\[
[H,X_i]=0,\qquad i=1,2,3,
\]
and which transforms as a vector under the total angular momentum
\[
\vec J=\vec L+\frac{\hbar}{2}(\vec\sigma_1+\vec\sigma_2).
\]
Thus
\[
[J_i,X_j]=i\hbar\varepsilon_{ijk}X_k.
\]

We first construct the most general first-order vector operator. The available
vectorial directions generated by
$\vec x$, $\vec p$, $\vec\sigma_1$, and $\vec\sigma_2$ are
\[\{
\vec{x},\quad
\vec{p},\quad
\vec{L},\quad
\vec{\sigma}_1,\quad
\vec{\sigma}_2,\quad
\vec{\sigma}_1\wedge\vec{x},\quad
\vec{\sigma}_2\wedge\vec{x},\quad
\vec{\sigma}_1\wedge\vec{p},\quad
\vec{\sigma}_2\wedge\vec{p},\quad
\vec{\sigma}_1\wedge\vec{\sigma}_2\}.
\]
Combining these directions with rotational scalar factors and retaining only
terms which are at most first order in the momenta, one obtains the following
twenty-eight vector structures:
\begin{align*}
&\vec{\mathcal V}_1=\vec{x},\quad
\vec{\mathcal V}_2=\vec{p},\quad
\vec{\mathcal V}_3=\vec{\sigma}_1\wedge\vec{x},\quad
\vec{\mathcal V}_4=\vec{\sigma}_2\wedge\vec{x},\quad
\vec{\mathcal V}_5=\vec{\sigma}_1\wedge\vec{p},\quad
\vec{\mathcal V}_6=\vec{\sigma}_2\wedge\vec{p},\quad
\vec{\mathcal V}_7=(\vec{x},\vec{p})\vec{x},\\
&
\vec{\mathcal V}_8=(\vec{\sigma}_1,\vec{\sigma}_2)\vec{x},\quad
\vec{\mathcal V}_9=(\vec{\sigma}_1,\vec{\sigma}_2)\vec{p},\quad
\vec{\mathcal V}_{10}=((\vec{\sigma}_1\wedge\vec{x}),\vec{p})\vec{x},\quad
\vec{\mathcal V}_{11}=((\vec{\sigma}_2\wedge\vec{x}),\vec{p})\vec{x},\\
&
\vec{\mathcal V}_{12}=((\vec{\sigma}_1\wedge\vec{x}),\vec{p})(\vec{\sigma}_2\wedge\vec{x}),\quad
\vec{\mathcal V}_{13}=((\vec{\sigma}_2\wedge\vec{x}),\vec{p})(\vec{\sigma}_1\wedge\vec{x}),\quad
\vec{\mathcal V}_{14}=(\vec{\sigma}_1,\vec{\sigma}_2)(\vec{x},\vec{p})\vec{x},\\
&
\vec{\mathcal V}_{15}=(\vec{\sigma}_1,\vec{x})(\vec{\sigma}_2,\vec{x})\vec{x},\quad
\vec{\mathcal V}_{16}=(\vec{\sigma}_1,\vec{x})(\vec{\sigma}_2,\vec{x})\vec{p},\quad
\vec{\mathcal V}_{17}=(\vec{\sigma}_1,\vec{x})(\vec{\sigma}_2,\vec{p})\vec{x},\quad
\vec{\mathcal V}_{18}=(\vec{\sigma}_2,\vec{x})(\vec{\sigma}_1,\vec{p})\vec{x},\\
&
\vec{\mathcal V}_{19}=(\vec{\sigma}_1,\vec{x})(\vec{\sigma}_2,\vec{x})(\vec{x},\vec{p})\vec{x},\quad
\vec{\mathcal V}_{20}=(\vec{\sigma}_1,\vec{x})\vec{\sigma}_2,\quad
\vec{\mathcal V}_{21}=(\vec{\sigma}_2,\vec{x})\vec{\sigma}_1,\quad
\vec{\mathcal V}_{22}=(\vec{\sigma}_1,\vec{x})\vec{L},\\
&
\vec{\mathcal V}_{23}=(\vec{\sigma}_2,\vec{x})\vec{L},\quad
\vec{\mathcal V}_{24}=-((\vec{\sigma}_1\wedge\vec{x}),\vec{\sigma}_2)\vec{L},\quad
\vec{\mathcal V}_{25}=(\vec{\sigma}_1,\vec{p})\vec{\sigma}_2,\quad
\vec{\mathcal V}_{26}=(\vec{\sigma}_2,\vec{p})\vec{\sigma}_1,\\
&
\vec{\mathcal V}_{27}=(\vec{\sigma}_1,\vec{x})(\vec{x},\vec{p})\vec{\sigma}_2,\quad
\vec{\mathcal V}_{28}=(\vec{\sigma}_2,\vec{x})(\vec{x},\vec{p})\vec{\sigma}_1.
\end{align*}
Therefore an arbitrary first-order vector operator is initially written as
\[
\vec X_V=\sum_{j=1}^{28}f_j(r)\vec{\mathcal V}_j,
\]
where the coefficient functions $f_j$ are real functions of $r$. Since the
factors containing $\vec x$, $\vec p$, and the radial functions do not
commute in general, the expression must be symmetrized in order to obtain a
Hermitian operator. After applying the standard symmetrization procedure term by
term \cite{YT}, the most general first-order vector operator takes the following form:
\begin{align*}
\vec{X}_V &= \left[ f_1 - \frac{i\hbar}{2r} f_2' - \frac{i\hbar}{2} r f_7' - 2i\hbar f_7 \right] \vec{x} \\
&\quad + \left[ f_8 - \frac{i\hbar}{2r} f_9' + \frac{i\hbar}{2} (f_{12} + f_{13} - r f_{14}' - 4 f_{14} - f_{17} - f_{18}) \right] (\vec{\sigma}_1, \vec{\sigma}_2) \vec{x} \\
&\quad + \left[ f_{15} - \frac{i\hbar}{2r} (f_{16}' + f_{17}' + f_{18}') - \frac{i\hbar}{2} r f_{19}' - 3i\hbar f_{19} \right] (\vec{\sigma}_1, \vec{x}) (\vec{\sigma}_2, \vec{x}) \vec{x} \\
&\quad + \left[ f_{21} - \frac{i\hbar}{2} (f_{12} + f_{16} + f_{18} + f_{24}) - \frac{i\hbar}{2r} f_{26}' - \frac{i\hbar}{2} r f_{28}' - 2i\hbar f_{28} \right] (\vec{\sigma}_2, \vec{x}) \vec{\sigma}_1 \\
&\quad + \left[ f_{20} - \frac{i\hbar}{2} (f_{13} + f_{16} + f_{17} - f_{24}) - \frac{i\hbar}{2r} f_{25}' - \frac{i\hbar}{2} r f_{27}' - 2i\hbar f_{27} \right] (\vec{\sigma}_1, \vec{x}) \vec{\sigma}_2 \\
&\quad + \left[ f_3 - \frac{i\hbar}{2r} f_5' - \frac{i\hbar}{2} f_{10} + \frac{i\hbar}{2} f_{22} \right] (\vec{\sigma}_1 \wedge \vec{x})  + \left[ f_4 - \frac{i\hbar}{2r} f_6' - \frac{i\hbar}{2} f_{11} + \frac{i\hbar}{2} f_{23} \right] (\vec{\sigma}_2 \wedge \vec{x}) \\
&\quad + \left[ f_2 + f_9 (\vec{\sigma}_1, \vec{\sigma}_2) + f_{16} (\vec{\sigma}_1, \vec{x}) (\vec{\sigma}_2, \vec{x}) \right] \vec{p}  + \left[ f_{22} (\vec{\sigma}_1, \vec{x}) + f_{23} (\vec{\sigma}_2, \vec{x}) -  f_{24} ((\vec{\sigma}_1 \wedge \vec{x}), \vec{\sigma}_2) \right] \vec{L} \\
&\quad + f_7 (\vec{x}, \vec{p}) \vec{x} + f_{14} (\vec{\sigma}_1, \vec{\sigma}_2) \vec{x} (\vec{x}, \vec{p}) + f_{19} (\vec{\sigma}_1, \vec{x}) (\vec{\sigma}_2, \vec{x}) \vec{x} (\vec{x}, \vec{p})  + f_{27} (\vec{\sigma}_1, \vec{x}) \vec{\sigma}_2 (\vec{x}, \vec{p}) \\
&\quad + f_{28} (\vec{\sigma}_2, \vec{x}) \vec{\sigma}_1 (\vec{x}, \vec{p}) + f_5 (\vec{\sigma}_1 \wedge \vec{p}) + f_6 (\vec{\sigma}_2 \wedge \vec{p}) + f_{10} ((\vec{\sigma}_1 \wedge \vec{x}), \vec{p}) \vec{x} + f_{11} ((\vec{\sigma}_2 \wedge \vec{x}), \vec{p}) \vec{x} \\
&\quad + f_{12} (\vec{\sigma}_2 \wedge \vec{x}) ((\vec{\sigma}_1 \wedge \vec{x}), \vec{p}) + f_{13} (\vec{\sigma}_1 \wedge \vec{x}) ((\vec{\sigma}_2 \wedge \vec{x}), \vec{p}) \\
&\quad + f_{17} \vec{x} (\vec{\sigma}_1, \vec{x}) (\vec{\sigma}_2, \vec{p}) + f_{18} \vec{x} (\vec{\sigma}_2, \vec{x}) (\vec{\sigma}_1, \vec{p})  + f_{25} \vec{\sigma}_2 (\vec{\sigma}_1, \vec{p}) + f_{26} \vec{\sigma}_1 (\vec{\sigma}_2, \vec{p}).
\end{align*}

The classification is obtained by substituting the general vector ansatz
given above into the commutation condition
\[
[H,\vec X_V]=0
\]
and equating to zero the coefficients of the resulting independent differential
operator structures. This gives a large overdetermined system of radial
determining equations for the coefficient functions $f_j(r)$ and for the
potentials $V_i(r)$. The determining equations obtained in this way are listed
in reduced form in Appendix~\ref{app:determining_equations}.

The solution of this system follows the same algorithmic elimination procedure
as in the scalar and pseudo-scalar classifications
\cite{TuncerYurdusen2025,TurkkanTuncerYurdusen2026}. Since the resulting
branch-by-branch computation is lengthy and largely repetitive, we do not
reproduce the full elimination in the main text. Instead, we state the complete
classification theorem. Each family listed below is obtained from the
determining equations and can be verified directly by substituting the
corresponding potentials and vector integrals into $[H,\vec X_V]=0$.

For brevity, whenever an integral of motion coincides with one already listed,
we refer to it by the same notation without repeating its explicit expression.
In each case, we list the admissible potentials and the corresponding
non-trivial vector integrals of motion. We do not claim that all listed
integrals are mutually algebraically independent, rather, they represent
distinct vector solutions of the determining equations.
\begin{theorem}
Let $r=\|\vec{x}\|$. All constants appearing below are real. The symbols
$\lambda,\lambda_1,\lambda_2$ and $\beta$ denote real constants, while
$\epsilon$ satisfies $\epsilon^2=1$. Then the spherically symmetric
Hamiltonians admitting non-trivial first-order vector integrals of motion are
precisely those corresponding to the families listed below.
\begin{enumerate}
    \item $V_4(r) = 0$, $V_5(r) = 0$,
\begin{align*}
V_0(r) &= V_0(r),\quad
V_1(r) = \frac{\hbar}{r^2}, \quad
V_2(r) = \frac{\hbar^2}{4r^2},\quad 
V_3(r) = -\frac{3\hbar^2}{4r^4}+\frac{\lambda}{r^2}+\frac{V_0(r)}{r^2}.
\end{align*}
This system admits five non-trivial first-order vector integrals of motion, namely
\begin{align*}
\vec X_1 &=
\vec{p}-(\vec{\sigma}_1,\vec{\sigma}_2)\vec{p}
\\[1ex]
\vec X_2 &=
\frac{\hbar}{2r}\vec{x}
+\frac{i\hbar}{2r}(\vec{\sigma}_1\wedge \vec{x})
+\frac{\hbar}{2r}(\vec{\sigma}_1,\vec{x})\vec{\sigma}_2
+\frac{1}{r}(\vec{\sigma}_1,\vec{x})\vec{L},
\\[1ex]
\vec X_3 &=
\frac{\hbar}{2r}\vec{x}
+\frac{i\hbar}{2r}(\vec{\sigma}_2\wedge \vec{x})
+\frac{\hbar}{2r}(\vec{\sigma}_2,\vec{x})\vec{\sigma}_1
+\frac{1}{r}(\vec{\sigma}_2,\vec{x})\vec{L},
\\[1ex]
\vec X_4 
&=-(\vec{\sigma}_1,\vec{\sigma}_2)\vec{p}
+\frac{1}{r^2}(\vec{\sigma}_1,\vec{x})(\vec{\sigma}_2,\vec{x})\vec{p}
-\frac{\hbar}{2r^2}(\vec{\sigma}_1\wedge\vec{x})
-\frac{\hbar}{2r^2}(\vec{\sigma}_2\wedge\vec{x}) \\
&\quad
+\frac{i\hbar}{r^4}(\vec{\sigma}_1,\vec{x})(\vec{\sigma}_2,\vec{x})\vec{x} 
-\frac{i\hbar}{2r^2}(\vec{\sigma}_1,\vec{x})\vec{\sigma}_2
-\frac{i\hbar}{2r^2}(\vec{\sigma}_2,\vec{x})\vec{\sigma}_1.
\\[1ex]
\vec X_5 &=
-\frac{3\hbar}{2r}(\vec{\sigma}_1\wedge \vec{x})
+\frac{3\hbar}{2r}(\vec{\sigma}_2\wedge \vec{x})
+\frac{3i\hbar}{2r}(\vec{\sigma}_1,\vec{x})\vec{\sigma}_2
-\frac{3i\hbar}{2r}(\vec{\sigma}_2,\vec{x})\vec{\sigma}_1
-\frac{3}{r}\bigl((\vec{\sigma}_1\wedge \vec{x}),\vec{\sigma}_2\bigr)\vec{L}.
\end{align*}
\item $V_4(r)=0$, $V_5(r)=0$,
\begin{align*}
V_0(r) &= V_0(r),\quad
V_1(r) = \frac{\hbar}{r^2}, \quad
V_2(r) = \frac{\hbar^2}{4r^2},\quad
V_3(r) = V_3(r).
\end{align*}
This system admits three non-trivial first-order integrals of motion, namely $\vec X_2$, $\vec X_3$, and $\vec X_5$.
\item $V_4(r)=0$, $V_5(r)=0$,
\begin{align*}
V_0(r) &= V_0(r),\quad
V_1(r) = \frac{\hbar}{r^2}, \quad
V_2(r) = \lambda_{1}+\frac{\hbar^2}{4r^2},\quad
V_3(r) = -\frac{3\hbar^2}{4r^4}+\frac{\lambda_{2}+\lambda_{1}}{r^2}+\frac{V_0(r)}{r^2}.
\end{align*}
This system admits three non-trivial first-order integrals of motion, namely $\vec X_1$, $\vec X_2+\vec X_3$, and $\vec X_4$.
\item $V_4(r)=0$, $V_5(r)=0$,
\begin{align*}
V_0(r) &= V_0(r),\quad
V_1(r) = \frac{\hbar}{r^2}, \quad
V_2(r) = V_2(r),\quad
V_3(r) = \frac{-\hbar^2+\lambda r^2+r^2\bigl(V_0(r)+V_2(r)\bigr)}{r^4}.
\end{align*}
This system admits two non-trivial first-order integrals of motion, namely $\vec X_1-2\vec X_4$ and $\vec X_2+\vec X_3$.
\item $V_4(r)=0$, $V_5(r)=0$,
\begin{align*}
V_0(r) &= V_0(r),\quad
V_1(r) = \frac{\hbar}{r^2}, \quad
V_2(r) = V_2(r),\quad
V_3(r) = \frac{\lambda+V_0(r)-3V_2(r)}{r^2}.
\end{align*}
This system admits two non-trivial first-order integrals of motion, namely $\vec X_1$ and $\vec X_2+\vec X_3$.
\item $V_4(r)=0$, $V_5(r)=0$,
\begin{align*}
V_0(r) &= V_0(r),\quad
V_1(r) = \frac{\hbar}{r^2}, \quad
V_2(r) = V_2(r),\quad
V_3(r) = V_3(r).
\end{align*}
This system admits one non-trivial first-order integral of motion, namely $\vec X_2+\vec X_3$.
\item $V_4(r)=0$, $V_5(r)=0$,
\begin{align*}
V_0(r) &= \frac{2\hbar^2}{r^2}+\frac{\alpha^2 r^2}{2}+\lambda,\quad
V_1(r) = \frac{2\hbar}{r^2}, \quad
V_2(r) = -\frac{\alpha\hbar}{4}+\frac{\hbar^2}{r^2},\quad
V_3(r) = -\frac{\hbar^2}{r^4}.
\end{align*}
This system admits two non-trivial first-order integrals of motion, namely
\begin{align*}
\vec X_6 &=
-\frac{i\hbar}{r^2}(\vec{\sigma}_1\wedge \vec{x})
+\frac{i\hbar}{r^2}(\vec{\sigma}_2\wedge \vec{x})
+(\vec{\sigma}_1\wedge \vec{p})
-(\vec{\sigma}_2\wedge \vec{p})
\\
&\quad
-\left(\alpha+\frac{\hbar}{r^2}\right)(\vec{\sigma}_1,\vec{x})\vec{\sigma}_2
+\left(\alpha+\frac{\hbar}{r^2}\right)(\vec{\sigma}_2,\vec{x})\vec{\sigma}_1
-\frac{2}{r^2}(\vec{\sigma}_1,\vec{x})\vec{L}
+\frac{2}{r^2}(\vec{\sigma}_2,\vec{x})\vec{L},
\\[1ex]
\vec X_7 &=
-\frac{\hbar+\alpha r^2}{2r^2}(\vec{\sigma}_1\wedge \vec{x})
+\frac{\hbar+\alpha r^2}{2r^2}(\vec{\sigma}_2\wedge \vec{x})
+\frac{i\hbar}{2r^2}(\vec{\sigma}_1,\vec{x})\vec{\sigma}_2
-\frac{i\hbar}{2r^2}(\vec{\sigma}_2,\vec{x})\vec{\sigma}_1
\\
&\quad
-\frac{1}{r^2}\bigl((\vec{\sigma}_1\wedge \vec{x}),\vec{\sigma}_2\bigr)\vec{L}
-\frac{1}{2}\vec{\sigma}_2(\vec{\sigma}_1,\vec{p})
+\frac{1}{2}\vec{\sigma}_1(\vec{\sigma}_2,\vec{p}).
\end{align*}
\item $V_4(r)=0$, $V_5(r)=0$,
\begin{align*}
V_0(r) &= V_0(r),\quad
V_1(r) = \frac{2\hbar}{r^2}, \quad
V_2(r) = \lambda+\frac{3\hbar^2}{r^2}-V_0(r),\quad
V_3(r) = -\frac{\hbar^2}{r^4}.
\end{align*}
This system admits three non-trivial first-order integrals of motion, namely
\begin{align*}
\vec X_8 &=
\vec{p}
+\frac{4\hbar}{3r^2}(\vec{\sigma}_1\wedge \vec{x})
+\frac{4\hbar}{3r^2}(\vec{\sigma}_2\wedge \vec{x})
+\frac{1}{3}(\vec{\sigma}_1,\vec{\sigma}_2)\vec{p},
\\[1ex]
\vec X_9 &=
-\frac{4\hbar}{r^2}\vec{x}
-\frac{i\hbar}{r^2}(\vec{\sigma}_1\wedge \vec{x})
-\frac{i\hbar}{r^2}(\vec{\sigma}_2\wedge \vec{x})
+(\vec{\sigma}_1\wedge \vec{p})
+(\vec{\sigma}_2\wedge \vec{p})
\\
&\quad
-\frac{2\hbar}{r^2}(\vec{\sigma}_1,\vec{\sigma}_2)\vec{x}
+\frac{4\hbar}{r^4}(\vec{\sigma}_1,\vec{x})(\vec{\sigma}_2,\vec{x})\vec{x}
-\frac{\hbar}{r^2}(\vec{\sigma}_1,\vec{x})\vec{\sigma}_2
-\frac{\hbar}{r^2}(\vec{\sigma}_2,\vec{x})\vec{\sigma}_1
\\
&\quad
-\frac{2}{r^2}(\vec{\sigma}_1,\vec{x})\vec{L}
-\frac{2}{r^2}(\vec{\sigma}_2,\vec{x})\vec{L},
\\[1ex]
\vec X_{10} &=
\frac{5\hbar}{3r^2}(\vec{\sigma}_1\wedge \vec{x})
+\frac{5\hbar}{3r^2}(\vec{\sigma}_2\wedge \vec{x})
-\frac{2i\hbar}{r^2}(\vec{\sigma}_1,\vec{\sigma}_2)\vec{x}
+\frac{2}{3}(\vec{\sigma}_1,\vec{\sigma}_2)\vec{p}
\\
&\quad
+\frac{12i\hbar}{r^4}(\vec{\sigma}_1,\vec{x})(\vec{\sigma}_2,\vec{x})\vec{x}
+\frac{4}{r^2}\vec{x}\,(\vec{\sigma}_1,\vec{x})(\vec{\sigma}_2,\vec{p})
-\frac{8}{r^4}(\vec{\sigma}_1,\vec{x})(\vec{\sigma}_2,\vec{x})\vec{x}\,(\vec{x},\vec{p})
\\
&\quad
-\frac{3i\hbar}{r^2}(\vec{\sigma}_1,\vec{x})\vec{\sigma}_2
-\frac{3i\hbar}{r^2}(\vec{\sigma}_2,\vec{x})\vec{\sigma}_1
+\frac{2}{r^2}\bigl((\vec{\sigma}_1\wedge \vec{x}),\vec{\sigma}_2\bigr)\vec{L}
+\vec{\sigma}_2(\vec{\sigma}_1,\vec{p})
\\
&\quad
-3\vec{\sigma}_1(\vec{\sigma}_2,\vec{p})
+\frac{4}{r^2}(\vec{\sigma}_2,\vec{x})\vec{\sigma}_1(\vec{x},\vec{p}).
\end{align*}
\item $V_4(r)=0$, $V_5(r)=0$,
\begin{equation*}
V_0(r) = \frac{\alpha^2 r^2}{2}+\lambda,\quad
V_1(r) = 0, \quad
V_2(r) = -\frac{\alpha\hbar}{4},\quad
V_3(r) = 0.
\end{equation*}
This system admits two non-trivial first-order integrals of motion, namely
\begin{align*}
\vec X_{11} &=
(\vec{\sigma}_1\wedge \vec{p})
-(\vec{\sigma}_2\wedge \vec{p})
-\alpha(\vec{\sigma}_1,\vec{x})\vec{\sigma}_2
+\alpha(\vec{\sigma}_2,\vec{x})\vec{\sigma}_1,
\\[1ex]
\vec X_{12} &=
-\frac{\alpha}{2}(\vec{\sigma}_1\wedge \vec{x})
+\frac{\alpha}{2}(\vec{\sigma}_2\wedge \vec{x})
-\frac{1}{2}\vec{\sigma}_2(\vec{\sigma}_1,\vec{p})
+\frac{1}{2}\vec{\sigma}_1(\vec{\sigma}_2,\vec{p}).
\end{align*}
\item $V_4(r)=0$, $V_5(r)=0$,
\begin{align*}
V_0(r) &= \lambda_{1},\quad
V_1(r) = 0, \quad
V_2(r) = \lambda_{2}-\lambda_{1},\quad
V_3(r) = 0.
\end{align*}
This system admits four non-trivial first-order integrals of motion, namely
\begin{align*}
\vec X_{13} &= \vec{p},
\\[1ex]
\vec X_{14} &= (\vec{\sigma}_1\wedge \vec{p})+(\vec{\sigma}_2\wedge \vec{p}),
\\[1ex]
\vec X_{15} &= \vec{\sigma}_2(\vec{\sigma}_1,\vec{p})+\vec{\sigma}_1(\vec{\sigma}_2,\vec{p}),
\\[1ex]
\vec X_{16} &= (\vec{\sigma}_1,\vec{\sigma}_2)\vec{p}.
\end{align*}
\item $V_4(r)=0$, $V_5(r)=0$,
\begin{align*}
V_0(r) &= V_0(r),\quad
V_1(r) = 0, \quad
V_2(r) = \lambda-V_0(r),\quad
V_3(r) = 0.
\end{align*}
This system admits three non-trivial first-order integrals of motion, namely $\vec X_{13}+\frac13 \vec X_{16}$, $\vec X_{14}$, and $\vec X_{15}-\frac23 \vec X_{16}$.
\item $V_4(r)=0$, $V_5(r)=0$,
\begin{align*}
V_0(r) &= \frac{2\hbar^2}{r^2}+\lambda_{1},\quad
V_1(r) = \frac{2\hbar}{r^2}, \quad
V_2(r) = \lambda_{2}+\frac{\hbar^2}{r^2}-\lambda_{1},\quad
V_3(r) = -\frac{\hbar^2}{r^4}.
\end{align*}
This system admits two non-trivial first-order integrals of motion, namely $\frac34 \vec X_8+\frac14 \vec X_{13}-\frac14 \vec X_{16}$ and $\frac34 \vec X_8-\frac34 \vec X_{13}+\frac34 \vec X_{16}$.
\item $V_4(r)=0$, $V_5(r)=0$,
\begin{align*}
V_0(r) &= \frac{2\alpha^2(1+\beta r^2)}{\beta}
+\frac{\alpha\hbar}{2+2\beta r^2}
+\frac{\hbar^2\bigl(8+15\beta r^2+6\beta^2 r^4+8\epsilon(1+\beta r^2)^{3/2}\bigr)}{8(r+\beta r^3)^2}
+\lambda,\\
V_1(r) &= \frac{\hbar}{r^2}\left(1+\dfrac{\epsilon}{\sqrt{1+\beta r^2}}\right),\\
V_2(r) &= \frac{\hbar\Bigl(4\alpha r^2(1+\beta r^2)+\hbar\bigl(4+4\epsilon\sqrt{1+\beta r^2}
+\beta r^2(7+2\beta r^2+4\epsilon\sqrt{1+\beta r^2})\bigr)\Bigr)}{8(r+\beta r^3)^2},\\
V_3(r) &= \frac{\hbar\Bigl(-2\hbar-3\beta\hbar r^2+4\alpha\beta r^4(1+\beta r^2)
-2\epsilon\hbar(1+\beta r^2)^{3/2}\Bigr)}{4r^4(1+\beta r^2)^2}.
\end{align*}
This system admits two non-trivial first-order integrals of motion, namely
\begin{align*}
\vec X_{17} &=
-\frac{i\hbar\bigl(\epsilon+\sqrt{1+\beta r^2}\bigr)}{2\epsilon r^2}
(\vec{\sigma}_1\wedge \vec{x})
+\frac{i\hbar\bigl(\epsilon+\sqrt{1+\beta r^2}\bigr)}{2\epsilon r^2}
(\vec{\sigma}_2\wedge \vec{x})
+(\vec{\sigma}_1\wedge \vec{p})
-(\vec{\sigma}_2\wedge \vec{p})
\\
&\quad
+\left(2\alpha+\frac{\beta\hbar}{2-2\epsilon\sqrt{1+\beta r^2}}\right)
(\vec{\sigma}_1,\vec{x})\vec{\sigma}_2 \\
&\quad
-\frac{\epsilon\bigl(\epsilon+\sqrt{1+\beta r^2}\bigr)\bigl(-\beta\hbar+4\alpha(-1+\epsilon\sqrt{1+\beta r^2})\bigr)}
{2\beta r^2}
(\vec{\sigma}_2,\vec{x})\vec{\sigma}_1
\\
&\quad
-\frac{\epsilon+\sqrt{1+\beta r^2}}{\epsilon r^2}
(\vec{\sigma}_1,\vec{x})\vec{L}
+\frac{\epsilon+\sqrt{1+\beta r^2}}{\epsilon r^2}
(\vec{\sigma}_2,\vec{x})\vec{L},
\\[1ex]
\vec X_{18} &=
\left(\alpha-\frac{\hbar\bigl(\epsilon+\sqrt{1+\beta r^2}\bigr)}{4\epsilon r^2}\right)
(\vec{\sigma}_1\wedge \vec{x})
+\left(-\alpha+\frac{\hbar\bigl(\epsilon+\sqrt{1+\beta r^2}\bigr)}{4\epsilon r^2}\right)
(\vec{\sigma}_2\wedge \vec{x})
\\
&\quad
+\frac{i\hbar\bigl(\epsilon+\sqrt{1+\beta r^2}\bigr)}{4\epsilon r^2}
(\vec{\sigma}_1,\vec{x})\vec{\sigma}_2
-\frac{i\hbar\bigl(\epsilon+\sqrt{1+\beta r^2}\bigr)}{4\epsilon r^2}
(\vec{\sigma}_2,\vec{x})\vec{\sigma}_1
\\
&\quad
-\frac{\epsilon+\sqrt{1+\beta r^2}}{2\epsilon r^2}
\bigl((\vec{\sigma}_1\wedge \vec{x}),\vec{\sigma}_2\bigr)\vec{L}
-\frac12 \vec{\sigma}_2(\vec{\sigma}_1,\vec{p})
+\frac12 \vec{\sigma}_1(\vec{\sigma}_2,\vec{p}).
\end{align*}
\item $V_4(r)=0$, $V_5(r)=0$,
\begin{align*}
V_0(r) &= \frac{\hbar^2\bigl(8+15\beta r^2+6\beta^2 r^4+8\epsilon(1+\beta r^2)^{3/2}\bigr)}{8(r+\beta r^3)^2}
+\lambda,\\
V_1(r) &= \frac{\hbar}{r^2}\left(1+\dfrac{\epsilon}{\sqrt{1+\beta r^2}}\right),\\
V_2(r) &= \frac{\hbar^2\bigl(4+4\epsilon\sqrt{1+\beta r^2}+\beta r^2(7+2\beta r^2+4\epsilon\sqrt{1+\beta r^2})\bigr)}{8(r+\beta r^3)^2},\\
V_3(r) &= -\frac{\hbar^2\bigl(2+3\beta r^2+2\epsilon(1+\beta r^2)^{3/2}\bigr)}{4r^4(1+\beta r^2)^2}.
\end{align*}
This system admits three non-trivial first-order integrals of motion, namely $\vec X_{13}-\vec X_{16}$, $\vec X_{17}\big|_{\alpha=0}$, and $\vec X_{18}\big|_{\alpha=0}$.
\item $V_4(r)=0$, $V_5(r)=0$,
\begin{align*}
V_0(r) &= \frac{\alpha^2 r^2}{2}
+\frac{\alpha\hbar}{4+4\beta r^2}
+\frac{\hbar^2\bigl(8(1+\beta r^2)^{3/2}
+\epsilon\bigl(8+3\beta r^2(5+2\beta r^2)\bigr)\bigr)}
{8\epsilon(r+\beta r^3)^2}
+\lambda,\\
V_1(r) &= \frac{\hbar}{r^2}\left(1+\dfrac{\epsilon}{\sqrt{1+\beta r^2}}\right),\\
V_2(r)
&=
\frac{\alpha\hbar}{4(1+\beta r^2)}
+
\frac{\hbar^2\left(
4+7\beta r^2+2\beta^2 r^4
+4\epsilon(1+\beta r^2)^{3/2}
\right)}
{8r^2(1+\beta r^2)^2},\\
V_3(r)
&=
\frac{\alpha\beta\hbar}{2(1+\beta r^2)}
-
\frac{\hbar^2\left(2+3\beta r^2+2\epsilon(1+\beta r^2)^{3/2}\right)}
{4r^4(1+\beta r^2)^2}.
\end{align*}
This system admits one non-trivial first-order integral of motion, namely
\[
\vec X_{19}=\vec X_{17}-\alpha\Bigl((\vec{\sigma}_1,\vec{x})\vec{\sigma}_2
-(\vec{\sigma}_2,\vec{x})\vec{\sigma}_1\Bigr).
\]
\item $V_4(r)=0$, $V_5(r)=0$,
\begin{align*}
V_0(r) &= \frac{\hbar^2\bigl(8(1+\beta r^2)^{3/2}
+\epsilon\bigl(8+3\beta r^2(5+2\beta r^2)\bigr)\bigr)}
{8\epsilon(r+\beta r^3)^2}
+\lambda,\\
V_1(r) &= \frac{\hbar}{r^2}\left(1+\dfrac{\epsilon}{\sqrt{1+\beta r^2}}\right),\\
V_2(r) &= \frac{\beta\hbar^2\bigl(1+3\epsilon\sqrt{1+\beta r^2}
+2\beta r^2(1+\epsilon\sqrt{1+\beta r^2})\bigr)}
{8(1+\beta r^2)^2\bigl(-1+\epsilon\sqrt{1+\beta r^2}\bigr)},\\
V_3(r)
&=
-\frac{\hbar^2\left(2+3\beta r^2
+2\epsilon(1+\beta r^2)^{3/2}\right)}
{4r^4(1+\beta r^2)^2}.
\end{align*}
This system admits two non-trivial first-order integrals of motion, namely $\vec X_{13}-\vec X_{16}$ and $\vec X_{17}\big|_{\alpha=0}$.
\item $V_4(r)=0$, $V_5(r)=0$,
\begin{align*}
V_0(r) &= V_0(r),\quad
V_1(r) = V_1(r), \quad
V_2(r) = V_2(r),\quad
V_3(r) = \frac{\lambda+V_0(r)-3V_2(r)}{r^2}.
\end{align*}
This system admits one non-trivial first-order integral of motion, namely $\vec X_1$.
\item $V_4(r)=0$,
\begin{align*}
V_0(r) = \frac{11\hbar^2}{4r^2}+\lambda,\quad
V_1(r) = \frac{7\hbar}{2r^2},\quad
V_2(r) = \frac{5\hbar^2}{4r^2}, \quad
V_3(r) = -\frac{\hbar^2}{r^4},\quad
V_5(r) = \frac{1}{2r^2}.
\end{align*}
This system admits six non-trivial first-order integrals of motion, namely
\begin{align*}
\vec X_{20} &=
-\frac{i\hbar}{2r}(\vec{\sigma}_1\wedge \vec{x})
+\frac{i\hbar}{2r}(\vec{\sigma}_2\wedge \vec{x})
+\frac{1}{r}\bigl((\vec{\sigma}_1\wedge \vec{x}),\vec{p}\bigr)\vec{x}
-\frac{1}{r}\bigl((\vec{\sigma}_2\wedge \vec{x}),\vec{p}\bigr)\vec{x}
\\
&\quad
+\frac{\hbar}{2r}(\vec{\sigma}_1,\vec{x})\vec{\sigma}_2
-\frac{\hbar}{2r}(\vec{\sigma}_2,\vec{x})\vec{\sigma}_1,
\\[1ex]
\vec X_{21} &=
-\frac{1}{2r}(\vec{\sigma}_1\wedge \vec{x})
+\frac{1}{2r}(\vec{\sigma}_2\wedge \vec{x})
-\frac{i}{2r}(\vec{\sigma}_1,\vec{x})\vec{\sigma}_2
+\frac{i}{2r}(\vec{\sigma}_2,\vec{x})\vec{\sigma}_1
\\
&\quad
-\frac{1}{\hbar r}\bigl((\vec{\sigma}_1\wedge \vec{x}),\vec{\sigma}_2\bigr)\vec{L}
-\frac{r}{\hbar}\vec{\sigma}_2(\vec{\sigma}_1,\vec{p})
+\frac{r}{\hbar}\vec{\sigma}_1(\vec{\sigma}_2,\vec{p})
\\
&\quad
+\frac{1}{\hbar r}(\vec{\sigma}_1,\vec{x})\vec{\sigma}_2(\vec{x},\vec{p})
-\frac{1}{\hbar r}(\vec{\sigma}_2,\vec{x})\vec{\sigma}_1(\vec{x},\vec{p}),
\\[1ex]
\vec X_{22} &=
\frac{\hbar}{4r}(\vec{\sigma}_1\wedge \vec{x})
+\frac{\hbar}{4r}(\vec{\sigma}_2\wedge \vec{x})
-\frac{i\hbar}{2r}(\vec{\sigma}_1,\vec{\sigma}_2)\vec{x}
+\frac{4i\hbar}{r^3}(\vec{\sigma}_1,\vec{x})(\vec{\sigma}_2,\vec{x})\vec{x}
\\
&\quad
+\frac{1}{r}(\vec{\sigma}_1,\vec{x})(\vec{\sigma}_2,\vec{x})\vec{p}
+\frac{1}{r}\vec{x}(\vec{\sigma}_2,\vec{x})(\vec{\sigma}_1,\vec{p})
-\frac{2}{r^3}(\vec{\sigma}_1,\vec{x})(\vec{\sigma}_2,\vec{x})\vec{x}(\vec{x},\vec{p})
\\
&\quad
-\frac{3i\hbar}{4r}(\vec{\sigma}_1,\vec{x})\vec{\sigma}_2
-\frac{3i\hbar}{4r}(\vec{\sigma}_2,\vec{x})\vec{\sigma}_1
-\frac{1}{2r}\bigl((\vec{\sigma}_1\wedge \vec{x}),\vec{\sigma}_2\bigr)\vec{L}
\\
&\quad
-\frac{r}{2}\vec{\sigma}_2(\vec{\sigma}_1,\vec{p})
+\frac{r}{2}\vec{\sigma}_1(\vec{\sigma}_2,\vec{p})
+\frac{1}{2r}(\vec{\sigma}_1,\vec{x})\vec{\sigma}_2(\vec{x},\vec{p})
-\frac{1}{2r}(\vec{\sigma}_2,\vec{x})\vec{\sigma}_1(\vec{x},\vec{p}),
\\[1ex]
\vec X_{23} &=
\frac{i\hbar}{r}\vec{x}
+r\vec{p}
-\frac{1}{r}\vec{x}(\vec{x},\vec{p})
-\frac{i\hbar}{r}(\vec{\sigma}_1,\vec{\sigma}_2)\vec{x}
-r(\vec{\sigma}_1,\vec{\sigma}_2)\vec{p}
+\frac{1}{r}(\vec{\sigma}_1,\vec{\sigma}_2)\vec{x}(\vec{x},\vec{p}),
\\[1ex]
\vec X_{24} &=
-\frac{i\hbar}{r^2}\vec{x}
+\frac{1}{r^2}\vec{x}(\vec{x},\vec{p})
+\frac{i\hbar}{r^2}(\vec{\sigma}_1,\vec{\sigma}_2)\vec{x}
-\frac{1}{r^2}(\vec{\sigma}_1,\vec{\sigma}_2)\vec{x}(\vec{x},\vec{p}),
\\[1ex]
\vec X_{25} &=
\frac{1}{r}\vec{x}
-\frac{1}{r}(\vec{\sigma}_1,\vec{\sigma}_2)\vec{x}.
\end{align*}
\item $V_4(r)=0$,
\begin{align*}
V_0(r) = V_0(r),\quad
V_1(r) = \frac{7\hbar}{2r^2},\quad
V_2(r) = \frac{5\hbar^2}{4r^2}, \quad
V_3(r) = -\frac{\hbar^2}{r^4},\quad
V_5(r) = \frac{1}{2r^2}.
\end{align*}
This system admits five non-trivial first-order integrals of motion, namely $\vec X_{20}$, $\vec X_{21}$, $\vec X_{22}$, $\vec X_{23}$, and $\vec X_{25}$.
\item $V_4(r)=0$,
\begin{align*}
V_0(r) &= V_0(r),\quad
V_1(r) = \frac{7\hbar}{2r^2},\quad
V_2(r) = \lambda+\frac{\hbar^2}{3r^2}+\frac{V_0(r)}{3},\\
V_3(r) &= -\frac{\hbar^2}{r^4},\quad
V_5(r) = \frac{1}{2r^2}.
\end{align*}
This system admits four non-trivial first-order integrals of motion, namely $\vec X_{22}$, $\vec X_{23}$, $\vec X_{24}$, and $\vec X_{25}$.
\item $V_4(r)=0$,
\begin{align*}
V_0(r) = V_0(r),\quad
V_1(r) = \frac{7\hbar}{2r^2},\quad
V_2(r) = V_2(r), \quad
V_3(r) = -\frac{\hbar^2}{r^4},\quad
V_5(r) = \frac{1}{2r^2}.
\end{align*}
This system admits three non-trivial first-order integrals of motion, namely $\vec X_{22}$, $\vec X_{23}$, and $\vec X_{25}$.
\item $V_4(r)=0$,
\begin{align*}
V_0(r) &= V_0(r),\quad
V_1(r) = \frac{7\hbar}{2r^2},\quad
V_2(r) = \lambda-\frac{3\hbar^2}{2r^2}+V_0(r),\\
V_3(r) &= \frac{9\hbar^2-4\lambda r^2-4r^2V_0(r)}{2r^4},\quad
V_5(r) = \frac{1}{2r^2}.
\end{align*}
This system admits five non-trivial first-order integrals of motion, namely $\vec X_{20}$, $\vec X_{21}$, $\vec X_{23}$, $\vec X_{24}$, and $\vec X_{25}$.
\item $V_4(r)=0$,
\begin{align*}
V_0(r) = V_0(r),\quad
V_1(r) = \frac{7\hbar}{2r^2},\quad
V_2(r) = V_2(r), \quad
V_3(r) = \frac{3\hbar^2}{2r^4}-\frac{2V_2(r)}{r^2},\quad
V_5(r) = \frac{1}{2r^2}.
\end{align*}
This system admits four non-trivial first-order integrals of motion, namely $\vec X_{20}$, $\vec X_{21}$, $\vec X_{23}$, and $\vec X_{25}$.
\item $V_4(r)=0$,
\begin{align*}
V_0(r) = V_0(r),\quad
V_1(r) = \frac{7\hbar}{2r^2},\quad
V_2(r) = V_2(r), \quad
V_3(r) = \frac{\lambda+V_0(r)-3V_2(r)}{r^2},\quad
V_5(r) = \frac{1}{2r^2}.
\end{align*}
This system admits three non-trivial first-order integrals of motion, namely $\vec X_{23}$, $\vec X_{24}$, and $\vec X_{25}$
\item $V_4(r)=0$,
\begin{align*}
V_0(r) = \frac{3\hbar^2}{4r^2}+\lambda,\quad
V_1(r) = \frac{3\hbar}{2r^2},\quad
V_2(r) = \frac{\hbar^2}{4r^2}, \quad
V_3(r) = 0,\quad
V_5(r) = \frac{1}{2r^2}.
\end{align*}
This system admits six non-trivial first-order integrals of motion, namely $\vec X_{20}$, $\vec X_{21}$, $\vec X_{23}$, $\vec X_{24}$, $\vec X_{25}$, and
\begin{align*}
\vec X_{26} 
&=
-\frac{\hbar}{4r}(\vec{\sigma}_1\wedge \vec{x})
-\frac{\hbar}{4r}(\vec{\sigma}_2\wedge \vec{x})
+\frac{i\hbar}{2r}(\vec{\sigma}_1,\vec{\sigma}_2)\vec{x}
\\
&\quad
+\frac{1}{r}(\vec{\sigma}_1,\vec{x})(\vec{\sigma}_2,\vec{x})\vec{p}
-\frac{1}{2r}\vec{x}(\vec{\sigma}_2,\vec{x})(\vec{\sigma}_1,\vec{p})
-\frac{1}{2r}\vec{x}(\vec{\sigma}_1,\vec{x})(\vec{\sigma}_2,\vec{p})
\\
&\quad
-\frac{i\hbar}{4r}(\vec{\sigma}_1,\vec{x})\vec{\sigma}_2
-\frac{i\hbar}{4r}(\vec{\sigma}_2,\vec{x})\vec{\sigma}_1.
\end{align*}
\item $V_4(r)=0$,
\begin{align*}
V_0(r) = V_0(r),\quad
V_1(r) = \frac{3\hbar}{2r^2},\quad
V_2(r) = \frac{\hbar^2}{4r^2}, \quad
V_3(r) = 0,\quad
V_5(r) = \frac{1}{2r^2}.
\end{align*}
This system admits five non-trivial first-order integrals of motion, namely $\vec X_{20}$, $\vec X_{21}$, $\vec X_{23}$, $\vec X_{25}$, and $\vec X_{26}$.
\item $V_4(r)=0$,
\begin{align*}
V_0(r) = V_0(r),\quad
V_1(r) = \frac{3\hbar}{2r^2},\quad
V_2(r) = \lambda+\frac{V_0(r)}{3}, \quad
V_3(r) = 0,\quad
V_5(r) = \frac{1}{2r^2}.
\end{align*}
This system admits four non-trivial first-order integrals of motion, namely $\vec X_{23}$, $\vec X_{24}$, $\vec X_{25}$, and $\vec X_{26}$.
\item $V_4(r)=0$,
\begin{align*}
V_0(r) = V_0(r),\quad
V_1(r) = \frac{3\hbar}{2r^2},\quad
V_2(r) = V_2(r), \quad
V_3(r) = 0,\quad
V_5(r) = \frac{1}{2r^2}.
\end{align*}
This system admits three non-trivial first-order integrals of motion, namely $\vec X_{23}$, $\vec X_{25}$, and $\vec X_{26}$.
\item $V_4(r)=0$,
\begin{align*}
V_0(r) &= V_0(r),\quad
V_1(r) = \frac{3\hbar}{2r^2},\quad
V_2(r) = \lambda-\frac{\hbar^2}{2r^2}+V_0(r),\\
V_3(r) &= \frac{3\hbar^2-4\lambda r^2-4r^2V_0(r)}{2r^4},\quad
V_5(r) = \frac{1}{2r^2}.
\end{align*}
This system admits five non-trivial first-order integrals of motion, namely $\vec X_{20}$, $\vec X_{21}$, $\vec X_{23}$, $\vec X_{24}$, and $\vec X_{25}$.
\item $V_4(r)=0$,
\begin{align*}
V_0(r) = V_0(r),\quad
V_1(r) = \frac{3\hbar}{2r^2},\quad
V_2(r) = V_2(r), \quad
V_3(r) = \frac{\hbar^2-4r^2V_2(r)}{2r^4},\quad
V_5(r) = \frac{1}{2r^2}.
\end{align*}
This system admits four non-trivial first-order integrals of motion, namely $\vec X_{20}$, $\vec X_{21}$, $\vec X_{23}$, and $\vec X_{25}$.
\item $V_4(r)=0$,
\begin{align*}
V_0(r) = V_0(r),\quad
V_1(r) = \frac{3\hbar}{2r^2},\quad
V_2(r) = V_2(r), \quad
V_3(r) = \frac{\lambda+V_0(r)-3V_2(r)}{r^2},\quad
V_5(r) = \frac{1}{2r^2}.
\end{align*}
This system admits three non-trivial first-order integrals of motion, namely $\vec X_{23}$, $\vec X_{24}$, and $\vec X_{25}$
\item $V_4(r)=0$,
\begin{align*}
V_0(r) &= V_0(r),\quad
V_1(r) = V_1(r),\quad
V_2(r) = \lambda+\frac{\hbar^2}{4r^2}+V_0(r)-\frac{1}{2}\hbar V_1(r),\\
V_3(r) &= -\frac{3\hbar^2+8\lambda r^2+8r^2V_0(r)-6\hbar r^2V_1(r)}{4r^4},\quad
V_5(r) = \frac{1}{2r^2}.
\end{align*}
This system admits five non-trivial first-order integrals of motion, namely $\vec X_{20}$, $\vec X_{21}$, $\vec X_{23}$, $\vec X_{24}$, and $\vec X_{25}$.
\item $V_4(r)=0$,
\begin{align*}
V_0(r) &= V_0(r),\quad
V_1(r) = V_1(r),\quad
V_2(r) = V_2(r),\\
V_3(r) &= -\frac{\hbar^2-2\hbar r^2V_1(r)+8r^2V_2(r)}{4r^4},\quad
V_5(r) = \frac{1}{2r^2}.
\end{align*}
This system admits four non-trivial first-order integrals of motion, namely $\vec X_{20}$, $\vec X_{21}$, $\vec X_{23}$, and $\vec X_{25}$.
\item $V_4(r)=0$,
\begin{align*}
V_0(r) = V_0(r),\quad
V_1(r) = V_1(r),\quad
V_2(r) = V_2(r), \quad
V_3(r) = \frac{\lambda+V_0(r)-3V_2(r)}{r^2},\quad
V_5(r) = \frac{1}{2r^2}.
\end{align*}
This system admits three non-trivial first-order integrals of motion, namely $\vec X_{23}$, $\vec X_{24}$, and $\vec X_{25}$.
\item $V_4(r)=0$,
\begin{align*}
V_0(r) = V_0(r),\quad
V_1(r) = V_1(r),\quad
V_2(r) = V_2(r), \quad
V_3(r) = V_3(r),\quad
V_5(r) = \frac{1}{2r^2}.
\end{align*}
This system admits two non-trivial first-order integrals of motion, namely $\vec X_{23}$ and $\vec X_{25}$.
\item $V_4(r)=0$,
\begin{align*}
V_0(r) = V_0(r),\quad
V_1(r) = -\frac{\hbar}{2r^2},\quad
V_2(r) = \lambda-V_0(r), \quad
V_3(r) = 0,\quad
V_5(r) = -\frac{1}{2r^2}.
\end{align*}
This system admits five non-trivial first-order integrals of motion, namely
\begin{align*}
\vec X_{27} &=
-\frac{i\hbar}{2r}(\vec{\sigma}_1\wedge \vec{x})
-\frac{i\hbar}{2r}(\vec{\sigma}_2\wedge \vec{x})
-\frac{\hbar}{r}(\vec{\sigma}_1,\vec{\sigma}_2)\vec{x}
+\frac{1}{r}\bigl((\vec{\sigma}_1\wedge \vec{x}),\vec{p}\bigr)\vec{x}
+\frac{1}{r}\bigl((\vec{\sigma}_2\wedge \vec{x}),\vec{p}\bigr)\vec{x}
\\
&\quad
+\frac{4\hbar}{r^3}(\vec{\sigma}_1,\vec{x})(\vec{\sigma}_2,\vec{x})\vec{x}
-\frac{\hbar}{2r}(\vec{\sigma}_1,\vec{x})\vec{\sigma}_2
-\frac{\hbar}{2r}(\vec{\sigma}_2,\vec{x})\vec{\sigma}_1,
\\[1ex]
\vec X_{28} &=
-\frac{i\hbar}{2r}\vec{x}
-\frac{r}{2}\vec{p}
+\frac{1}{2r}\vec{x}(\vec{x},\vec{p})
-\frac{3i\hbar}{2r}(\vec{\sigma}_1,\vec{\sigma}_2)\vec{x}
-\frac{r}{2}(\vec{\sigma}_1,\vec{\sigma}_2)\vec{p}
+\frac{1}{2r}(\vec{\sigma}_1,\vec{\sigma}_2)\vec{x}(\vec{x},\vec{p})
\\
&\quad
+\frac{6i\hbar}{r^3}(\vec{\sigma}_1,\vec{x})(\vec{\sigma}_2,\vec{x})\vec{x}
+\frac{1}{r}(\vec{\sigma}_1,\vec{x})(\vec{\sigma}_2,\vec{x})\vec{p}
+\frac{2}{r}\vec{x}(\vec{\sigma}_2,\vec{x})(\vec{\sigma}_1,\vec{p})
-\frac{3}{r^3}(\vec{\sigma}_1,\vec{x})(\vec{\sigma}_2,\vec{x})\vec{x}(\vec{x},\vec{p})
\\
&\quad
-\frac{i\hbar}{r}(\vec{\sigma}_1,\vec{x})\vec{\sigma}_2
-\frac{i\hbar}{r}(\vec{\sigma}_2,\vec{x})\vec{\sigma}_1
-\frac{1}{r}\bigl((\vec{\sigma}_1\wedge \vec{x}),\vec{\sigma}_2\bigr)\vec{L}
-r\,\vec{\sigma}_2(\vec{\sigma}_1,\vec{p})
+r\,\vec{\sigma}_1(\vec{\sigma}_2,\vec{p})
\\
&\quad
+\frac{1}{r}(\vec{\sigma}_1,\vec{x})\vec{\sigma}_2(\vec{x},\vec{p})
-\frac{1}{r}(\vec{\sigma}_2,\vec{x})\vec{\sigma}_1(\vec{x},\vec{p}),
\\[1ex]
\vec X_{29} &=
\frac{1}{r}(\vec{\sigma}_1\wedge \vec{x})
+\frac{1}{r}(\vec{\sigma}_2\wedge \vec{x})
-\frac{2i}{r}(\vec{\sigma}_1,\vec{\sigma}_2)\vec{x}
+\frac{8i}{r^3}(\vec{\sigma}_1,\vec{x})(\vec{\sigma}_2,\vec{x})\vec{x}
+\frac{4}{\hbar r}\vec{x}(\vec{\sigma}_2,\vec{x})(\vec{\sigma}_1,\vec{p})
\\
&\quad
-\frac{4}{\hbar r^3}(\vec{\sigma}_1,\vec{x})(\vec{\sigma}_2,\vec{x})\vec{x}(\vec{x},\vec{p})
-\frac{i}{r}(\vec{\sigma}_1,\vec{x})\vec{\sigma}_2
-\frac{i}{r}(\vec{\sigma}_2,\vec{x})\vec{\sigma}_1
-\frac{2}{\hbar r}\bigl((\vec{\sigma}_1\wedge \vec{x}),\vec{\sigma}_2\bigr)\vec{L}
\\
&\quad
-\frac{2r}{\hbar}\vec{\sigma}_2(\vec{\sigma}_1,\vec{p})
+\frac{2r}{\hbar}\vec{\sigma}_1(\vec{\sigma}_2,\vec{p})
+\frac{2}{\hbar r}(\vec{\sigma}_1,\vec{x})\vec{\sigma}_2(\vec{x},\vec{p})
-\frac{2}{\hbar r}(\vec{\sigma}_2,\vec{x})\vec{\sigma}_1(\vec{x},\vec{p}),
\\[1ex]
\vec X_{30} &=
-\frac{i\hbar}{r^2}\vec{x}
+\frac{1}{r^2}\vec{x}(\vec{x},\vec{p})
-\frac{i\hbar}{r^2}(\vec{\sigma}_1,\vec{\sigma}_2)\vec{x}
+\frac{1}{r^2}(\vec{\sigma}_1,\vec{\sigma}_2)\vec{x}(\vec{x},\vec{p})
\\
&\quad
+\frac{2i\hbar}{r^4}(\vec{\sigma}_1,\vec{x})(\vec{\sigma}_2,\vec{x})\vec{x}
-\frac{2}{r^4}(\vec{\sigma}_1,\vec{x})(\vec{\sigma}_2,\vec{x})\vec{x}(\vec{x},\vec{p}),
\\[1ex]
\vec X_{31} &=
\frac{1}{r}\vec{x}
+\frac{1}{r}(\vec{\sigma}_1,\vec{\sigma}_2)\vec{x}
-\frac{2}{r^3}(\vec{\sigma}_1,\vec{x})(\vec{\sigma}_2,\vec{x})\vec{x}.
\end{align*}
\item $V_4(r)=0$,
\begin{align*}
V_0(r) = V_0(r),\quad
V_1(r) = -\frac{\hbar}{2r^2},\quad
V_2(r) = V_2(r), \quad
V_3(r) = 0,\quad
V_5(r) = -\frac{1}{2r^2}.
\end{align*}
This system admits four non-trivial first-order integrals of motion, namely $\vec X_{27}$, $\vec X_{28}$, $\vec X_{29}$, and $\vec X_{31}$.
\item $V_4(r)=0$,
\begin{align*}
V_0(r) = V_0(r),\quad
V_1(r) = -\frac{\hbar}{2r^2},\quad
V_2(r) = V_2(r), \quad
V_3(r) = \frac{\lambda+V_0(r)+V_2(r)}{r^2},\quad
V_5(r) = -\frac{1}{2r^2}.
\end{align*}
This system admits three non-trivial first-order integrals of motion, namely $\vec X_{28}-\frac{\hbar}{2}\vec X_{29}$, $\vec X_{30}$, and $\vec X_{31}$.
\item $V_4(r)=0$,
\begin{align*}
V_0(r) = V_0(r),\quad
V_1(r) = -\frac{\hbar}{2r^2},\quad
V_2(r) = V_2(r), \quad
V_3(r) = V_3(r),\quad
V_5(r) = -\frac{1}{2r^2}.
\end{align*}
This system admits two non-trivial first-order integrals of motion, namely $\vec X_{28}-\frac{\hbar}{2}\vec X_{29}$ and $\vec X_{31}$.
\end{enumerate}
\end{theorem}

\section{Symmetry Algebras}

In this section we describe the polynomial symmetry algebras associated with two
representative systems from the classification theorem. The notation used in the
classification is retained: $\vec X_a$ denotes a vector integral of motion, and we
write
\[
\vec X_a=(X_{a,1},X_{a,2},X_{a,3})
\]
for its components. The total angular momentum is denoted by $J_i$, and
\[
K=\vec{\sigma}_1\cdot\vec{\sigma}_2
\]
is the spin-exchange scalar. We also use
\[
[A,B]=AB-BA,
\qquad
\{A,B\}=AB+BA .
\]
Repeated indices are summed over $1,2,3$.

\subsection{Symmetry Algebra of Case 2}

We first consider the family
\[
V_4(r)=V_5(r)=0,
\]
\[
V_0(r)=V_0(r),
\qquad
V_1(r)=\frac{\hbar}{r^2},
\qquad
V_2(r)=\frac{\hbar^2}{4r^2},
\qquad
V_3(r)=V_3(r),
\]
where $V_0(r)$ and $V_3(r)$ are arbitrary radial functions. According to the
classification theorem, this system admits the three non-trivial first-order
vector integrals given by
\[
\vec X_2
=
\frac{\hbar}{2r}\vec{x}
+\frac{i\hbar}{2r}(\vec{\sigma}_1\wedge \vec{x})
+\frac{\hbar}{2r}(\vec{\sigma}_1, \vec{x})\vec{\sigma}_2
+\frac{1}{r}(\vec{\sigma}_1, \vec{x})\vec{L},
\]
\[
\vec X_3
=
\frac{\hbar}{2r}\vec{x}
+\frac{i\hbar}{2r}(\vec{\sigma}_2\wedge \vec{x})
+\frac{\hbar}{2r}(\vec{\sigma}_2, \vec{x})\vec{\sigma}_1
+\frac{1}{r}(\vec{\sigma}_2, \vec{x})\vec{L},
\]
and
\[
\vec X_5
=-\frac{3\hbar}{2r}(\vec{\sigma}_1\wedge \vec{x})
+\frac{3\hbar}{2r}(\vec{\sigma}_2\wedge \vec{x})
+\frac{3i\hbar}{2r}(\vec{\sigma}_1, \vec{x})\vec{\sigma}_2
-\frac{3i\hbar}{2r}(\vec{\sigma}_2, \vec{x})\vec{\sigma}_1
-\frac{3}{r}\bigl((\vec{\sigma}_1\wedge \vec{x}), \vec{\sigma}_2\bigr)\vec{L}.
\]

For the algebraic closure it is convenient to introduce the following linear
combinations of their components:
\[
P_i=X_{2,i}+X_{3,i},
\]
and
\[
U_i=X_{2,i}-X_{3,i}+\frac{i}{3}X_{5,i},
\qquad
V_i=X_{2,i}-X_{3,i}-\frac{i}{3}X_{5,i}.
\]
This case also admits the following scalar integral of motion \cite{TuncerYurdusen2025}
\[
\mathcal Y
=
\frac{
(\vec{\sigma}_1,\vec x)
(\vec{\sigma}_2,\vec x)
}{r^2}.
\]
Note that the operators 
\[ K=(\vec{\sigma}_1,\vec{\sigma}_2),\qquad \mathcal Y=(\vec{\sigma}_1,\hat{\vec x})(\vec{\sigma}_2,\hat{\vec x}), \qquad \hat{\vec x}=\frac{\vec x}{r}, \] 
satisfy 
\[ K^2=3-2K,\qquad \mathcal Y^2=1,\qquad (1-K)(1+\mathcal Y)=0. \] 
The first identity follows from the fact that $K$ has eigenvalue $-3$ on the singlet sector and eigenvalue $1$ on the triplet sector. The second identity follows immediately from 
\[ (\vec{\sigma}_1,\hat{\vec x})^2=(\vec{\sigma}_2,\hat{\vec x})^2=1. \]
The last identity follows from the singlet projector 
\[ \Pi_0=\frac{1-K}{4}. \] 
Since $(\vec{\sigma}_1+\vec{\sigma}_2)\Pi_0=0$, one has 
\[ (\vec{\sigma}_2,\hat{\vec x})\Pi_0=-(\vec{\sigma}_1,\hat{\vec x})\Pi_0, \] 
and hence $\mathcal Y \Pi_0=-\Pi_0$. This is equivalent to 
\[ (1-K)(1+\mathcal Y)=0. \]

\begin{proposition}
For the above system, the associative algebra
\[
\mathcal A_2
=
\langle H,K,\mathcal Y,J_i,P_i,U_i,V_i\mid i=1,2,3\rangle
\]
is polynomially closed. 
$\mathcal Y$ is central in $\mathcal A_2$:
\[
[\mathcal Y,H]=[\mathcal Y,K]=[\mathcal Y,J_i]
=
[\mathcal Y,P_i]=[\mathcal Y,U_i]=[\mathcal Y,V_i]=0.
\]
The action of $K$ on the vector generators is
\[
[K,P_i]=0,
\qquad
[K,U_i]=4U_i,
\qquad
[K,V_i]=-4V_i.
\]
The angular momentum generators satisfy
\[
[J_i,J_j]=i\hbar\varepsilon_{ijk}J_k,
\]
and the generators $P_i,U_i,V_i$ transform as vectors:
\[
[J_i,P_j]=i\hbar\varepsilon_{ijk}P_k,
\qquad
[J_i,U_j]=i\hbar\varepsilon_{ijk}U_k,
\qquad
[J_i,V_j]=i\hbar\varepsilon_{ijk}V_k.
\]
The non-vanishing commutators among $P_i,U_i,V_i$ are
\[
[P_i,P_j]
=
2i\hbar\,\varepsilon_{ijk}(1+\mathcal Y)J_k,
\]
and
\[
[U_i,V_j]
=
4(1-K\mathcal Y)\{J_i,J_j\}
+
4i\hbar\,\varepsilon_{ijk}(1-\mathcal Y)J_k.
\]
The remaining commutators among $P_i,U_i,V_i$ vanish:
\[
[P_i,U_j]=0,
\qquad
[P_i,V_j]=0,
\qquad
[U_i,U_j]=0,
\qquad
[V_i,V_j]=0.
\]
\end{proposition}

The vector sector contains the quadratic central element
\[
\mathcal C_P
=
\vec P^{\,2}+2(1+\mathcal Y)\vec J^{\,2},
\qquad
\vec P^{\,2}=\sum_{i=1}^3P_i^2,
\qquad
\vec J^{\,2}=\sum_{i=1}^3J_i^2.
\]

\paragraph{Algebraic interpretation.}
The central scalar $\mathcal Y$ allows one to decompose the algebra into
central quotients. In the quotient
\[
\mathcal A_2^{(+)}
=
\mathcal A_2/(\mathcal Y-1),
\]
the relation $(1-K)(1+\mathcal Y)=0$ implies $K=1$. Hence the generators
$U_i$ and $V_i$ vanish in this quotient, since
\[
[K,U_i]=4U_i,
\qquad
[K,V_i]=-4V_i.
\]
The remaining non-trivial vector generators are $J_i$ and $P_i$, satisfying
\[
[J_i,J_j]=i\hbar\varepsilon_{ijk}J_k,
\qquad
[J_i,P_j]=i\hbar\varepsilon_{ijk}P_k,
\qquad
[P_i,P_j]=4i\hbar\varepsilon_{ijk}J_k.
\]
After the rescaling $\widetilde P_i=P_i/2$, this becomes the standard
$\mathfrak{so}(4)$-type algebra. Equivalently, defining
\[
M_i=\frac12(J_i+\widetilde P_i),
\qquad
N_i=\frac12(J_i-\widetilde P_i),
\]
one obtains two commuting copies of $\mathfrak{so}(3)$:
\[
[M_i,M_j]=i\hbar\varepsilon_{ijk}M_k,
\qquad
[N_i,N_j]=i\hbar\varepsilon_{ijk}N_k,
\qquad
[M_i,N_j]=0.
\]

On the other hand, in the quotient
\[
\mathcal A_2^{(-)}
=
\mathcal A_2/(\mathcal Y+1),
\]
one has
\[
[P_i,P_j]=0.
\]
Thus the subalgebra generated by $J_i$ and $P_i$ is of
$\mathfrak e(3)$-type:
\[
[J_i,J_j]=i\hbar\varepsilon_{ijk}J_k,
\qquad
[J_i,P_j]=i\hbar\varepsilon_{ijk}P_k,
\qquad
[P_i,P_j]=0.
\]
Therefore the $J_i,P_i$ sector interpolates, through the central value of
$\mathcal Y$, between an $\mathfrak{so}(4)$-type algebra and an
$\mathfrak e(3)$-type algebra. The additional generators $U_i$ and $V_i$
complete this sector to the full polynomial algebra.

\subsection{Symmetry Algebra of Case 9}

We now consider the oscillator-type case
\[
V_4(r)=V_5(r)=0,
\]
\[
V_0(r)=\frac{\alpha^2r^2}{2}+C_1,
\qquad
V_1(r)=0,
\qquad
V_2(r)=-\frac{\alpha\hbar}{4},
\qquad
V_3(r)=0.
\]
Equivalently, the Hamiltonian takes the form
\[
H=
\frac{\vec p^{\,2}}{2}
+
\frac{\alpha^2r^2}{2}
+
C_1
-
\frac{\alpha\hbar}{4}K .
\]
This system admits the two non-trivial first-order vector integrals
\begin{align*}
\vec X_{11} &=
(\vec{\sigma}_1\wedge \vec{p})
-(\vec{\sigma}_2\wedge \vec{p})
-\alpha(\vec{\sigma}_1,\vec{x})\vec{\sigma}_2
+\alpha(\vec{\sigma}_2,\vec{x})\vec{\sigma}_1,
\\[1ex]
\vec X_{12} &=
-\frac{\alpha}{2}(\vec{\sigma}_1\wedge \vec{x})
+\frac{\alpha}{2}(\vec{\sigma}_2\wedge \vec{x})
-\frac{1}{2}\vec{\sigma}_2(\vec{\sigma}_1,\vec{p})
+\frac{1}{2}\vec{\sigma}_1(\vec{\sigma}_2,\vec{p}).
\end{align*}

We introduce the combinations
\[
U_i=X_{11,i}+2iX_{12,i},
\qquad
V_i=X_{11,i}-2iX_{12,i}.
\]
They are again integrals of motion
\[
[H,U_i]=[H,V_i]=0.
\]
Since $X_{11,i}$ and $X_{12,i}$ are Hermitian, $
U_i^\dagger=V_i$. The action of $K$ is diagonal
\[
[K,U_i]=-4U_i,
\qquad
[K,V_i]=4V_i.
\]

We already have $
K^2=3-2K.$
We therefore introduce the singlet and triplet projectors
\[
\Pi_0=\frac{1-K}{4},
\qquad
\Pi_1=\frac{K+3}{4}.
\]
They satisfy
$
\Pi_0^2=\Pi_0,\,
\Pi_1^2=\Pi_1,\,
\Pi_0\Pi_1=\Pi_1\Pi_0=0,\,
\Pi_0+\Pi_1=1.$ Using $K^2=3-2K$, the commutation relations with $K$ imply 
\[
KU_i=-3U_i,
\quad
U_iK=U_i, \quad KV_i=V_i, \quad V_iK=-3V_i.
\]
Equivalently,
\[
U_i=\Pi_0U_i\Pi_1,
\qquad
V_i=\Pi_1V_i\Pi_0.
\]
Thus $U_i$ maps the triplet sector into the singlet sector, whereas $V_i$
maps the singlet sector into the triplet sector. In particular,
\[
U_iU_j=0,
\qquad
V_iV_j=0
\]
for all $i,j$.

We define the ordered bilinear operators
\[
B_{ij}=U_iV_j,
\qquad
C_{ij}=V_jU_i.
\]
They satisfy
\[
B_{ij}=\Pi_0B_{ij}\Pi_0,
\qquad
C_{ij}=\Pi_1C_{ij}\Pi_1.
\]
Consequently,
\[
B_{ij}C_{kl}=0,
\qquad
C_{kl}B_{ij}=0.
\]
The adjoint structure is
\[
B_{ij}^\dagger=B_{ji},
\qquad
C_{ij}^\dagger=C_{ji}.
\]

It is also convenient to introduce
\[
F_{ij}=\frac{i}{4}[U_i,V_j],
\qquad
G_{ij}=\frac12\{U_i,V_j\}.
\]
Then
\[
B_{ij}=G_{ij}-2iF_{ij},
\qquad
C_{ij}=G_{ij}+2iF_{ij},
\]
or equivalently,
\[
F_{ij}=\frac{i}{4}(B_{ij}-C_{ij}),
\qquad
G_{ij}=\frac12(B_{ij}+C_{ij}).
\]

Now we provide the following proposition.
\begin{proposition}
For the oscillator-type Hamiltonian above, the associative symmetry algebra
\[
\mathcal A_9
=
\left\langle
H,K,J_i,U_i,V_i,B_{ij},C_{ij}
\;\middle|\;
i,j=1,2,3
\right\rangle
\]
is polynomially closed, where
$
B_{ij}=U_iV_j,
\quad
C_{ij}=V_jU_i
$
are the ordered quadratic composite generators.

The action of $K$ on the generators is
\[
[K,J_i]=0,
\quad
[K,U_i]=-4U_i,
\quad
[K,V_i]=4V_i, \quad [K,B_{ij}]=[K,C_{ij}]=0.
\]
The angular momentum generators satisfy
\[
[J_i,J_j]
=
i\hbar\varepsilon_{ijk}J_k.
\]
The generators $U_i$ and $V_i$ transform as vectors:
\[
[J_\ell,U_i]
=
i\hbar\varepsilon_{\ell im}U_m,
\qquad
[J_\ell,V_i]
=
i\hbar\varepsilon_{\ell im}V_m.
\]
The quadratic generators transform as rank-two tensor operators:
\[
[J_\ell,B_{ij}]
=
i\hbar
\left(
\varepsilon_{\ell im}B_{mj}
+
\varepsilon_{\ell jm}B_{im}
\right),
\]
\[
[J_\ell,C_{ij}]
=
i\hbar
\left(
\varepsilon_{\ell im}C_{mj}
+
\varepsilon_{\ell jm}C_{im}
\right).
\]
The commutators among the first-order vector generators are
\[
[U_i,U_j]=0,
\quad
[V_i,V_j]=0, \quad [U_i,V_j]
=
B_{ij}-C_{ij}.
\]
The commutators of the quadratic generators with $U_k$ and $V_k$ are
\[
[B_{ij},U_k]
=
U_iC_{kj},
\qquad
[C_{ij},U_k]
=
-B_{kj}U_i,
\]
\[
[B_{ij},V_k]
=
-C_{ik}V_j,
\qquad
[C_{ij},V_k]
=
V_jB_{ik}.
\]
Finally, the commutators among the quadratic generators are
\[
[B_{ij},B_{kl}]
=
U_iC_{kj}V_l
-
U_kC_{il}V_j,
\]
\[
[C_{ij},C_{kl}]
=
V_jB_{il}U_k
-
V_lB_{kj}U_i,
\]
\[
[B_{ij},C_{kl}]=0.
\]
\end{proposition}

The Hamiltonian appears in an internal trace relation. Define
\[
G_{\mathrm{tr}}=\sum_{i=1}^3G_{ii},
\qquad
F_{\mathrm{tr}}=\sum_{i=1}^3F_{ii}.
\]
Equivalently,
\[
G_{\mathrm{tr}}
=
\frac12\sum_{i=1}^3(B_{ii}+C_{ii}),
\qquad
F_{\mathrm{tr}}
=
\frac{i}{4}\sum_{i=1}^3(B_{ii}-C_{ii}).
\]
A direct computation gives
\[
F_{\mathrm{tr}}
+
\frac{i}{2}G_{\mathrm{tr}}
=
i(1-K)
\left(
8H+6\alpha\hbar-8C_1
\right).
\]
On the other hand,
\[
F_{\mathrm{tr}}
+
\frac{i}{2}G_{\mathrm{tr}}
=
\frac{i}{2}\sum_{i=1}^3B_{ii}.
\]
Therefore, defining
$
B^{(0)}=\sum_{i=1}^3B_{ii},
$
the trace relation is equivalently written as
\[
B^{(0)}
=
2(1-K)
\left(
8H+6\alpha\hbar-8C_1
\right).
\]
Since $1-K=4\Pi_0$, this can also be written as
\[
B^{(0)}
=
8\Pi_0
\left(
8H+6\alpha\hbar-8C_1
\right).
\]

\paragraph{Algebraic interpretation.}
The oscillator-type system gives a polynomial symmetry algebra rather than a
finite-dimensional Lie algebra. Indeed, the first-order vector generators
$U_i$ and $V_i$ do not close linearly under commutation, since
\[
[U_i,V_j]=B_{ij}-C_{ij},
\qquad
B_{ij}=U_iV_j,
\qquad
C_{ij}=V_jU_i.
\]
Thus their mixed commutators produce ordered quadratic composite operators.

The projector relations
\[
U_i=\Pi_0U_i\Pi_1,
\qquad
V_i=\Pi_1V_i\Pi_0
\]
show that $U_i$ and $V_i$ connect the triplet and singlet sectors in
opposite directions. Consequently, $B_{ij}$ and $C_{ij}$ are block-diagonal
operators supported on the singlet and triplet sectors, respectively. This
block structure explains both the nilpotency relations of the first-order
generators and the vanishing mixed commutators
\[
[B_{ij},C_{kl}]=0.
\]

Since $U_i$ and $V_i$ transform as vectors under rotations, the quadratic
operators $B_{ij}$ and $C_{ij}$ transform as rank-two tensor operators.
Each quadratic sector therefore decomposes covariantly according to
\[
\mathbf 3\otimes\mathbf 3
=
\mathbf 1\oplus\mathbf 3\oplus\mathbf 5.
\]
For $B_{ij}$, this decomposition is
\[
B_{ij}
=
\frac{1}{3}\delta_{ij}B^{(0)}
+
\varepsilon_{ijk}B^{(1)}_k
+
B^{(2)}_{ij},
\]
where
\[
B^{(0)}=\sum_{i=1}^3B_{ii},
\qquad
B^{(1)}_k=\frac{1}{2}\varepsilon_{kij}B_{ij},
\]
and
\[
B^{(2)}_{ij}
=
\frac{1}{2}(B_{ij}+B_{ji})
-\frac{1}{3}\delta_{ij}B^{(0)}.
\]
The same decomposition applies to $C_{ij}$.

The trace relation
\[
B^{(0)}
=
2(1-K)
\left(
8H+6\alpha\hbar-8C_1
\right)
\]
shows that the scalar component of the $B$-sector is not independent, but is
determined by the Hamiltonian and the spin scalar $K$. Since
$
1-K=4\Pi_0,
$
the appearance of this factor is consistent with the fact that $B_{ij}$ is
supported entirely on the singlet sector.

Thus the algebra possesses a natural singlet--triplet block structure: the
operators $H$, $K$, $J_i$, $B_{ij}$, and $C_{ij}$ preserve the two
spin sectors, whereas $U_i$ and $V_i$ intertwine them.

\section{Exact solution via scalar reduction of a first-order vector integral}

In this section we study a special exactly solvable member of the family \eqref{eq:general_H_full}. Our aim is to use the additional first-order vector integral in the same spirit as the eigenvalue constructions appearing in the superintegrability literature: rather than treating the vector integral componentwise, we construct from it a scalar operator commuting with the Hamiltonian and the standard rotational invariants, and then solve the resulting common eigenvalue problem explicitly.

We consider the special choice
\begin{equation}\label{eq:special_choice_exact_revised}
V_2(r)=V_3(r)=V_4(r)=V_5(r)=0,
\qquad
V_1(r)=\frac{\hbar}{r^2}.
\end{equation}
Then the Hamiltonian takes the form
\begin{equation}\label{eq:H_special_exact_revised}
H=-\frac{\hbar^2}{2}\Delta+V_0(r)+\frac{\hbar}{2r^2}(\vec{\sigma}_1+\vec{\sigma}_2,\vec{L}).
\end{equation}
Introducing the total spin operator
\begin{equation}\label{eq:S_total_exact_revised}
\vec S=\frac{1}{2}(\vec{\sigma}_1+\vec{\sigma}_2),
\end{equation}
we may rewrite \eqref{eq:H_special_exact_revised} as
\begin{equation}\label{eq:H_LS_exact_revised}
{
H=-\frac{\hbar^2}{2}\Delta+V_0(r)+\frac{\hbar}{r^2}\,\vec L\cdot \vec S.
}
\end{equation}

The total angular momentum
\begin{equation}\label{eq:J_total_exact_revised}
\vec J=\vec L+\hbar\vec S
\end{equation}
satisfies
\begin{equation}\label{eq:J_comm_exact_revised}
[H,\vec J]=0,
\qquad
[H,\vec J^{\,2}]=0,
\qquad
[H,J_3]=0.
\end{equation}

At the outset one should keep in mind that the wave function of the system is not a two-component Pauli spinor, but rather a four-component spinor taking values in the tensor product space
\[
\mathbb C^2\otimes \mathbb C^2 \cong \mathbb C^4.
\]
Equivalently, in the uncoupled spin basis
\[
\{|\uparrow\uparrow\rangle,\ |\uparrow\downarrow\rangle,\ |\downarrow\uparrow\rangle,\ |\downarrow\downarrow\rangle\},
\]
one may write
\[
\Psi(\vec x)=
\begin{pmatrix}
\psi_{++}(\vec x)\\
\psi_{+-}(\vec x)\\
\psi_{-+}(\vec x)\\
\psi_{--}(\vec x)
\end{pmatrix}.
\]
Under addition of angular momenta, the spin space decomposes as
\begin{equation}\label{eq:spin_decomp_exact_revised}
\frac12\otimes\frac12=0\oplus1,
\end{equation}
that is, into a one-dimensional singlet subspace and a three-dimensional triplet subspace. The Hamiltonian preserves this decomposition, and the exact solution may therefore be constructed separately in the singlet and triplet sectors.

For the choice \eqref{eq:special_choice_exact_revised}, the system admits the first-order vector integral
\[
\vec X=\vec X_2+\vec X_3.
\]
Rather than selecting a particular component of the vector integral
$\vec X$, we form a rotational scalar that is compatible with the
commuting operators $H$, $\vec J^{\,2}$, and $J_3$.
Accordingly, we introduce
\begin{equation}\label{eq:K_def_exact_revised}
{
\mathcal Q=\vec J\cdot \vec X.
}
\end{equation}
Note that the operator $\mathcal Q$ is Hermitian.
This is the natural analogue, in the present setting, of the scalar combinations used in eigenvalue constructions for vector integrals of motion.

A direct computation shows that $\vec X$ simplifies drastically:
\begin{equation}\label{eq:X_simplified_exact_revised}
{
\vec X=(\vec{\sigma}_1+\vec{\sigma}_2)\cdot \hat{\vec x}\;\vec J
=
2(\vec S\cdot \hat{\vec x})\,\vec J,
\qquad
\hat{\vec x}=\frac{\vec x}{r}.
}
\end{equation}
Consequently,
\begin{equation}\label{eq:K_simplified_exact_revised}
{
\mathcal Q=2(\vec S\cdot \hat{\vec x})\,\vec J^{\,2}.
}
\end{equation}
Since $\vec X$ is an integral of motion and both $\vec J$ and $\vec X$ transform as vector operators, the scalar operator $\mathcal Q=\vec J\cdot\vec X$ satisfies
\[
[H,\mathcal Q]=0,
\qquad
[\vec J^{\,2},\mathcal Q]=0,
\qquad
[J_3,\mathcal Q]=0.
\]
Therefore, the set
\begin{equation}\label{eq:commuting_set_exact_revised}
{
\{H,\;\vec J^{\,2},\;J_3,\;\mathcal Q\}
}
\end{equation}
is a commuting family.

We now turn to the common eigenvalue problem
\begin{equation}\label{eq:common_eigenvalue_problem_exact_revised}
H\Psi=E\Psi,
\qquad
\vec J^{\,2}\Psi=\hbar^2 j(j+1)\Psi,
\qquad
J_3\Psi=\hbar m\Psi,
\qquad
\mathcal Q\Psi=\kappa\Psi.
\end{equation}
The discussion naturally splits into the singlet and triplet sectors.

In the singlet sector $s=0$, one has $\vec S=0$, and therefore $\vec X=0$ and $\mathcal Q=0$ identically. Thus the singlet sector is governed simply by a scalar central-potential problem. The singlet spinor is
\begin{equation}\label{eq:singlet_spinor_exact_revised}
\chi_{\mathrm{singlet}}
=
\frac{1}{\sqrt2}
\begin{pmatrix}
0\\
1\\
-1\\
0
\end{pmatrix},
\end{equation}
and the corresponding wave functions may be written as
\begin{equation}\label{eq:singlet_ansatz_exact_revised}
\Psi^{(0)}_{njm}(r,\theta,\phi)
=
R^{(0)}_{nj}(r)\,
Y_{jm}(\theta,\phi)\,
\chi_{\mathrm{singlet}}.
\end{equation}
Substitution into the Schrödinger equation gives
\begin{equation}\label{eq:radial_singlet_exact_revised}
-\frac{\hbar^2}{2}\left(R^{(0)\prime\prime}+\frac{2}{r}R^{(0)\prime}\right)
+
\left(
V_0(r)+\frac{\hbar^2}{2r^2}j(j+1)
\right)R^{(0)}
=
ER^{(0)}.
\end{equation}
Thus the singlet sector reduces to the standard scalar central potential problem (since $\vec S=0$ implies $\vec L\cdot\vec S=0$ identically) and does not involve the additional integral $\mathcal Q$.

\medskip

Before proceeding further, it is instructive to analyze explicitly how the additional operator $\mathcal Q$ acts on the four-component spinor in the uncoupled basis. Since $\vec J^{\,2}$ is fixed on a common eigenspace, the eigenvalue equation $\mathcal Q\Psi=\kappa\Psi$ reduces to
\begin{equation}
(\vec S\cdot \hat{\vec x})\Psi=\nu\Psi,
\qquad \kappa=2\hbar^2 j(j+1)\nu.
\end{equation}

Using
\[
\vec S=\frac{1}{2}(\vec\sigma_1+\vec\sigma_2)
=\frac{1}{2}(\vec\sigma\otimes I_2+I_2\otimes\vec\sigma),
\]
one finds that $\vec S\cdot\hat{\vec x}$ is represented by the $4\times4$ matrix
\[
\vec S\cdot\hat{\vec x}
=
\begin{pmatrix}
\cos\theta & \frac12 e^{-i\phi}\sin\theta & \frac12 e^{-i\phi}\sin\theta & 0\\
\frac12 e^{i\phi}\sin\theta & 0 & 0 & \frac12 e^{-i\phi}\sin\theta\\
\frac12 e^{i\phi}\sin\theta & 0 & 0 & \frac12 e^{-i\phi}\sin\theta\\
0 & \frac12 e^{i\phi}\sin\theta & \frac12 e^{i\phi}\sin\theta & -\cos\theta
\end{pmatrix}.
\]

Therefore the equation $(\vec S\cdot\hat{\vec x})\Psi=\nu\Psi$ yields a system of coupled equations for the components $\psi_{++},\psi_{+-},\psi_{-+},\psi_{--}$. In particular, subtracting the second and third equations gives
\begin{equation}
\nu(\psi_{+-}-\psi_{-+})=0.
\end{equation}

Hence, for $\nu=\pm1$, one obtains the constraint
\begin{equation}
{
\psi_{+-}=\psi_{-+},
}
\end{equation}
showing explicitly that the operator $\mathcal Q$ imposes nontrivial linear relations among the components of the four-component spinor.

Since $\vec S=0$ in the singlet sector, the operator $\mathcal Q$ vanishes identically there. Thus its nontrivial action appears entirely in the triplet sector. Since $\vec J^{\,2}$ is fixed on a common eigenspace, equation \eqref{eq:K_simplified_exact_revised} shows that the $\mathcal Q$-eigenvalue problem reduces to the eigenvalue problem for $\vec S\cdot \hat{\vec x}$. Restricted to the triplet sector, $\vec S\cdot \hat{\vec x}$ is the projection of a spin-1 operator onto the radial direction and therefore has the three eigenvalues
\begin{equation}\label{eq:helicity_values_exact_revised}
\nu,
\qquad
\nu=-1,0,1.
\end{equation}
Hence the corresponding eigenvalues of $\mathcal Q$ are
\begin{equation}\label{eq:kappa_values_exact_revised}
{
\kappa_\nu=2\hbar^2 j(j+1)\nu,
\qquad
\nu=-1,0,1.
}
\end{equation}

It is therefore natural to work in a local helicity basis. Let
\[
\{|1,1\rangle,\ |1,0\rangle,\ |1,-1\rangle\}
\]
denote the standard eigenbasis of $S_z$. Passing to the coupled basis $\mathbb{C}^2 \otimes \mathbb{C}^2 = \mathbb{C}^1 \oplus \mathbb{C}^3$, we identify the triplet sector with the spin-1 representation spanned by $\{|1,1\rangle,|1,0\rangle,|1,-1\rangle\}$. In this basis, the helicity eigenvectors $\chi_\nu$ are represented as three-component vectors. A convenient choice of normalized eigenvectors of $\vec S\cdot \hat{\vec x}$ is
\begin{equation}\label{eq:chi_plus_exact_revised}
\chi_{+1}(\theta,\phi)=
\begin{pmatrix}
e^{-i\phi}\cos^2\!\frac{\theta}{2}\\[1mm]
\dfrac{1}{\sqrt2}\sin\theta\\[1mm]
e^{i\phi}\sin^2\!\frac{\theta}{2}
\end{pmatrix},
\end{equation}
\begin{equation}\label{eq:chi_zero_exact_revised}
\chi_0(\theta,\phi)=
\begin{pmatrix}
-\dfrac{e^{-i\phi}}{\sqrt2}\sin\theta\\[1mm]
\cos\theta\\[1mm]
\dfrac{e^{i\phi}}{\sqrt2}\sin\theta
\end{pmatrix},
\end{equation}
\begin{equation}\label{eq:chi_minus_exact_revised}
\chi_{-1}(\theta,\phi)=
\begin{pmatrix}
e^{-i\phi}\sin^2\!\frac{\theta}{2}\\[1mm]
-\dfrac{1}{\sqrt2}\sin\theta\\[1mm]
e^{i\phi}\cos^2\!\frac{\theta}{2}
\end{pmatrix},
\end{equation}
which satisfy
\begin{equation}\label{eq:helicity_eigen_eq_exact_revised}
(\vec S\cdot \hat{\vec x})\chi_\nu=\nu\,\chi_\nu,
\qquad
\nu=-1,0,1.
\end{equation}

Since the operators $\vec J^{\,2}$, $J_3$ and $\mathcal Q$ act on the angular and spin variables, it is natural to separate the radial dependence from a combined spin-angular part. Accordingly, in the triplet sector we look for solutions of \eqref{eq:common_eigenvalue_problem_exact_revised} in the form
\begin{equation}\label{eq:ansatz_exact_revised}
\Psi_{njm\nu}(r,\theta,\phi)
=
R_{nj}(r)\,
e^{im\phi}\,
F_{jm}^{(\nu)}(\theta)\,
\chi_\nu(\theta,\phi),
\qquad
\nu=-1,0,1.
\end{equation}
This ansatz makes the role of the fourth quantum number $\nu$ completely explicit: it labels the three helicity sectors selected by the scalar operator $\mathcal Q$. Since the orbital angular momentum $\ell$ is a non-negative integer and the total spin 
is $s=1$ in the triplet sector, the total angular momentum quantum number $j$ is also 
a non-negative integer, and $m$ takes the values
\begin{equation}\label{eq:quantum_numbers_exact_revised}
j = 0, 1, 2, \ldots, \qquad 
m = -j, -j+1, \ldots, j-1, j.
\end{equation}
Note that the case $j=0$ is not included in the Coulomb and harmonic oscillator solutions considered below, since it does not admit a radial solution that is regular at the origin within the standard bound-state domain considered here.

Substituting \eqref{eq:ansatz_exact_revised} into the eigenvalue equations for $\vec J^{\,2}$, $J_3$, and $\mathcal Q$, 
one finds that the angular function $F=F_{jm}^{(\nu)}(\theta)$ must satisfy
\begin{equation}\label{eq:angular_equation_exact_revised}
F_{\theta\theta}+\cot\theta\,F_\theta
+
\left(
j(j+1)
-\frac{m^2+\nu^2-2m\nu\cos\theta}{\sin^2\theta}
\right)F=0.
\end{equation}
Setting $z=\cos\theta$, equation \eqref{eq:angular_equation_exact_revised} becomes
\begin{equation}\label{eq:angular_jacobi_exact_revised}
(1-z^2)F'' - 2zF'
+
\left(
j(j+1)
-\frac{m^2+\nu^2-2m\nu z}{1-z^2}
\right)F=0.
\end{equation}
A Frobenius analysis at the singular points $z=\pm 1$ shows that the indicial exponents are
$\tfrac{|m-\nu|}{2}$ at $z=+1$ and $\tfrac{|m+\nu|}{2}$ at $z=-1$.
Accordingly, we factor out the singular behaviour by writing
\begin{equation}\label{eq:F_factored_exact_revised}
F(z)=(1-z)^{\frac{|m-\nu|}{2}}(1+z)^{\frac{|m+\nu|}{2}}G(z).
\end{equation}
Substituting \eqref{eq:F_factored_exact_revised} into \eqref{eq:angular_jacobi_exact_revised}, one finds that $G$ satisfies the Jacobi differential equation
\begin{equation}\label{eq:G_jacobi_exact_revised}
(1-z^2)G''+\bigl[(\beta-\alpha)-(\alpha+\beta+2)z\bigr]G'
+\bigl[j(j+1)-\tfrac{(\alpha+\beta)(\alpha+\beta+2)}{4}\bigr]G=0,
\end{equation}
with
\begin{equation}\label{eq:alpha_beta_exact_revised}
\alpha=|m-\nu|,\qquad \beta=|m+\nu|.
\end{equation}
The regular solution on $[-1,1]$ is $G=P_{n_\theta}^{(\alpha,\beta)}(z)$, where $n_\theta$ is a
non-negative integer determined by the eigenvalue condition
\begin{equation}\label{eq:n_condition_exact_revised}
j(j+1)=\Bigl(n_\theta+\tfrac{\alpha+\beta}{2}\Bigr)\Bigl(n_\theta+\tfrac{\alpha+\beta}{2}+1\Bigr),
\end{equation}
which gives
\begin{equation}\label{eq:n_value_exact_revised}
n_\theta
=
j-\frac{|m-\nu|+|m+\nu|}{2}
=
j-\max(|m|,|\nu|).
\end{equation}
The requirement $n_\theta\geq 0$ yields the admissibility condition
\begin{equation}\label{eq:admissibility_exact_revised}
j\geq\max(|m|,|\nu|).
\end{equation}
Hence the regular solution of \eqref{eq:angular_equation_exact_revised} valid for all 
values of $m$ is
\begin{equation}\label{eq:angular_solution_exact_revised}
F_{jm}^{(\nu)}(\theta)
=
N_{jm\nu}\,
(1-z)^{\frac{|m-\nu|}{2}}
(1+z)^{\frac{|m+\nu|}{2}}
P_{j-\max(|m|,|\nu|)}^{(|m-\nu|,\;|m+\nu|)}(z),
\qquad z=\cos\theta,
\end{equation}
where $N_{jm\nu}$ is a normalization constant. The operator $\mathcal Q$ acts on the 
angular and spin degrees of freedom and selects the helicity sectors $\nu=-1,0,1$. 
The helicity label $\nu$ resolves the spin-angular part of the wave
function, but it does not enter the radial equation.

Using the identity
\[
\vec J^{\,2}=\vec L^{\,2}+\hbar^2\vec S^{\,2}+2\hbar\,\vec L\cdot\vec S,
\]
the angular part of the Hamiltonian may be rewritten as
\[
\frac{\vec L^{\,2}}{2r^2}+\frac{\hbar}{r^2}\vec L\cdot\vec S
=
\frac{\vec J^{\,2}-\hbar^2\vec S^{\,2}}{2r^2}.
\]
Hence, in the triplet sector $s=1$, the effective centrifugal term is
\[
\frac{\hbar^2}{2r^2}\bigl(j(j+1)-2\bigr).
\]
Substituting \eqref{eq:ansatz_exact_revised} into the Schrödinger equation $H\Psi=E\Psi$, we obtain the radial equation
\begin{equation}\label{eq:radial_equation_exact_revised}
-\frac{\hbar^2}{2}\left(R''+\frac{2}{r}R'\right)
+
\left(
V_0(r)+\frac{\hbar^2}{2r^2}(j(j+1)-2)
\right)R
=
ER.
\end{equation}
The effect of the spin-orbit term is therefore to replace the usual centrifugal parameter by $j(j+1)-2$ in the triplet sector.

We first consider the Coulomb potential
\begin{equation}\label{eq:Coulomb_potential_exact_revised}
V_0(r)=\frac{\mu}{r},
\qquad
\mu<0.
\end{equation}
In this case \eqref{eq:radial_equation_exact_revised} becomes
\begin{equation}\label{eq:radial_Coulomb_exact_revised}
R''+\frac{2}{r}R'
+
\left(
\frac{2E}{\hbar^2}-\frac{2\mu}{\hbar^2 r}-\frac{j(j+1)-2}{r^2}
\right)R=0.
\end{equation}
For bound states $E<0$, we set
\begin{equation}\label{eq:w_sigma_exact_revised}
w=\frac{\sqrt{-2E}}{\hbar},
\qquad
\sigma=2\frac{\sqrt{-2E}}{\hbar}=2w,
\end{equation}
and seek solutions of the form
\begin{equation}\label{eq:Coulomb_ansatz_exact_revised}
R(r)=e^{-wr}r^pL(\sigma r).
\end{equation}
Substituting this ansatz into \eqref{eq:radial_Coulomb_exact_revised} and requiring cancellation of the $r^{-2}$-term yields
\begin{equation}\label{eq:p_equation_exact_revised}
p(p+1)=j(j+1)-2.
\end{equation}
The regular root is
\begin{equation}\label{eq:p_value_exact_revised}
{
p=-\frac12+\sqrt{j^2+j-\frac74}.
}
\end{equation}
The remaining equation for $L$ is the associated Laguerre equation, whose regular solutions are
\begin{equation}\label{eq:Laguerre_solution_exact_revised}
L(\sigma r)=L_n^{(2p+1)}(\sigma r),
\qquad n=0,1,2,\dots.
\end{equation}
The quantization condition is
\begin{equation}\label{eq:quantization_exact_revised}
\frac{\mu}{\hbar\sqrt{-2E}}=-(n+p+1),
\end{equation}
hence the bound-state energies are
\begin{equation}\label{eq:Coulomb_energy_exact_revised}
E_{nj}
=
-\frac{\mu^2}{
2\hbar^2\left(
n+\frac12+\sqrt{j^2+j-\frac74}
\right)^2
}.
\end{equation}
Therefore the Coulomb radial functions in the triplet sector are
\begin{equation}\label{eq:Coulomb_radial_final_exact_revised}
{
R_{nj}(r)
=
\mathcal N_{nj}\,
e^{-wr}r^p
L_n^{(2p+1)}(2wr),
}
\end{equation}
with $p$ given by \eqref{eq:p_value_exact_revised}.

We next consider the isotropic harmonic oscillator potential
\begin{equation}\label{eq:oscillator_potential_exact_revised}
V_0(r)=\frac12\omega^2 r^2,
\qquad
\omega>0.
\end{equation}
Then \eqref{eq:radial_equation_exact_revised} becomes
\begin{equation}\label{eq:radial_oscillator_exact_revised}
R''+\frac{2}{r}R'
+
\left(
\frac{2E}{\hbar^2}
-
\frac{\omega^2}{\hbar^2}r^2
-
\frac{j(j+1)-2}{r^2}
\right)R=0.
\end{equation}
We use the standard oscillator ansatz
\begin{equation}\label{eq:oscillator_ansatz_exact_revised}
R(r)=e^{-\frac{\omega}{2\hbar} r^2}\,r^p\,L\!\left(\frac{\omega}{\hbar}r^2\right),
\end{equation}
where $p$ is again given by \eqref{eq:p_value_exact_revised}. Setting
\[
\rho=\frac{\omega}{\hbar}r^2,
\]
one finds that $L(\rho)$ satisfies the associated Laguerre equation. The regular solutions are
\begin{equation}\label{eq:oscillator_Laguerre_exact_revised}
L(\rho)=L_n^{\left(p+\frac12\right)}(\rho),
\qquad n=0,1,2,\dots.
\end{equation}
The corresponding bound-state energies are
\begin{equation}\label{eq:oscillator_energy_exact_revised}
{
E_{nj}
=
\hbar\omega\left(2n+p+\frac32\right)
=
\hbar\omega\left(
2n+1+\sqrt{j^2+j-\frac74}
\right).
}
\end{equation}
Hence the oscillator radial functions in the triplet sector are
\begin{equation}\label{eq:oscillator_radial_final_exact_revised}
{
R_{nj}(r)
=
\mathcal N'_{nj}\,
e^{-\frac{\omega r^2}{2\hbar}}r^p
L_n^{\left(p+\frac12\right)}(\frac{\omega}{\hbar} r^2).
}
\end{equation}

We may now write the full exact eigenfunctions. In the singlet sector they are of the form \eqref{eq:singlet_ansatz_exact_revised}, with radial equation \eqref{eq:radial_singlet_exact_revised}. In the triplet Coulomb case they are
\begin{equation}\label{eq:full_Coulomb_solution_exact_revised}
{
\Psi_{njm\nu}(r,\theta,\phi)
=
\mathcal C_{njm\nu}\,
e^{-wr}r^p\,
L_n^{(2p+1)}(2wr)\,
e^{im\phi}
F_{jm}^{(\nu)}(\theta)\,
\chi_\nu(\theta,\phi),
\qquad
\nu=-1,0,1,
}
\end{equation}
where $F_{jm}^{(\nu)}$ is given by \eqref{eq:angular_solution_exact_revised}, $\chi_\nu$ by \eqref{eq:chi_plus_exact_revised}--\eqref{eq:chi_minus_exact_revised}, and the energy by \eqref{eq:Coulomb_energy_exact_revised}. Here $\mathcal C_{njm\nu}$ is the normalization constant for the complete Coulomb eigenfunction. For the oscillator problem they are
\begin{equation}\label{eq:full_oscillator_solution_exact_revised}
{
\Psi_{njm\nu}(r,\theta,\phi)
=
\mathcal D_{njm\nu}\,
e^{-\frac{\omega r^2}{2\hbar}}r^p\,
L_n^{\left(p+\frac12\right)}(\frac{\omega}{\hbar} r^2)\,
e^{im\phi}
F_{jm}^{(\nu)}(\theta)\,
\chi_\nu(\theta,\phi),
\qquad
\nu=-1,0,1.
}
\end{equation}
Here $\mathcal D_{njm\nu}$ is the normalization constant for the complete
oscillator eigenfunction, and the corresponding energy is given by
\eqref{eq:oscillator_energy_exact_revised}.

We conclude that the additional first-order vector integral leads, through the
scalar operator
\[
\mathcal Q=\vec J\cdot\vec X,
\]
to an additional quantum number $\nu$, which labels the three helicity
sectors in the triplet spin sector. The angular problem is solved explicitly in
terms of Jacobi polynomials, while the radial problem is solved in terms of
associated Laguerre polynomials for both the Coulomb and harmonic oscillator
choices of $V_0(r)$. The radial equation is independent of $\nu$.
Therefore the energy eigenvalues depend only on the radial quantum number
$n$ and on $j$, but not on the helicity label. For fixed $n$ and $j$,
the levels are $(2j+1)$-fold degenerate with respect to
$m=-j,\ldots,j$ and threefold degenerate with respect to
$\nu=-1,0,1$. Hence each such energy level in the triplet sector has degeneracy
\[
3(2j+1),\qquad j\geq1.
\]
Thus, for this representative family, the additional vector integral provides a
complete separation of the spin-angular variables and leads to explicit
Coulomb- and oscillator-type bound-state solutions.

\section{Conclusion}

In this paper, we classified spherically symmetric Hamiltonians for two
interacting spin-$\frac12$ particles that admit non-trivial first-order vector
integrals of motion. The analysis was carried out for the class $V_4(r)=0$.
The Hamiltonian still contains the central, spin-orbit, spin-spin, tensor, and
quadratic spin-orbit interactions, with coefficients depending on the relative
distance.

We started from the most general Hermitian first-order vector operator formed
from the relative position, momentum, orbital angular momentum, and the two
spin vectors. The commutation condition with the Hamiltonian produced an
overdetermined system of radial determining equations. Solving this system led
to a complete list of $39$ Hamiltonian families admitting first-order vector
integrals within the class considered. Several families contain arbitrary
radial functions, while others correspond to special oscillator-type or
singular potentials. The classification also includes systems with a
non-vanishing quadratic spin-orbit interaction $V_5(r)$. Gauge-induced
families were identified separately from the genuinely spin-dependent systems.

These results extend our previous classifications of first-order scalar and
pseudo-scalar integrals for two interacting spin-$\frac12$ particles. The
vector case is richer because each integral has three components and may
generate further algebraic structures through its commutation relations. The
classification therefore provides not only new integrals of motion, but also a
large set of systems whose symmetry algebras and spectra can be studied in more
detail.

We also examined the symmetry algebras of two representative families. In one
case, the vector integrals lead to a polynomially closed algebra containing
$\mathfrak{so}(4)$- and $\mathfrak e(3)$-type reductions. In the
other case, the first-order generators produce quadratic tensor
operators and generate a polynomially closed algebra. These examples show that the
vector integrals found in the classification may generate rich algebraic
structures beyond ordinary Lie symmetry algebras.

For one representative family, we also used a scalar operator constructed from
a vector integral to solve the corresponding eigenvalue problem. The
spin-angular and radial variables separate, and the solutions are obtained in
terms of Jacobi and associated Laguerre polynomials. Explicit bound-state
energies and eigenfunctions were found for both the Coulomb and harmonic
oscillator potentials, together with a threefold helicity degeneracy in the
triplet sector.

The present work completes the vector part of the first-order classification
program for the Hamiltonian class $V_4(r)=0$. The next natural step is the
classification of first-order axial-vector integrals. Other possible
extensions include the full off-shell problem with $V_4(r)\neq0$, higher-order
integrals, and a systematic study of the representations and spectral
consequences of the polynomial algebras found here. These problems may lead to
further exactly solvable two-spin systems and to symmetry structures that do
not occur in spinless models.

\appendix

\section{Determining equations for first-order vector integrals}
\label{app:determining_equations}

In this appendix we list the determining equations obtained from the condition
\[
[H,\vec X]=0
\]
for the general first-order vector ansatz used in Section~3. The equations are
obtained after separating the commutator with respect to independent
differential operator structures. Repeated equations, equations differing only
by an overall sign, and equations with common nonzero factors are omitted. The
displayed system is not reduced to a minimal independent basis.

The classification theorem in Section~3 is obtained by solving this
overdetermined system by the same elimination procedure used in
\cite{TuncerYurdusen2025,TurkkanTuncerYurdusen2026}.

\subsection*{Third-order determining equations}

After imposing $V_4=0$, the third-order part of the determining system reduces
to the following equations:
\begin{align}
&V_5 f_{22}=0,\qquad
V_5 f_{23}=0, \label{eq:det3a}\\
&V_5(r^2f_{22}+f_5)=0,\qquad
V_5(r^2f_{23}+f_6)=0, \label{eq:det3b}\\
&V_5(f_{13}-3f_{24}+f_{27})=0,\qquad
V_5(f_{12}+3f_{24}+f_{28})=0, \label{eq:det3c}\\
&V_5\bigl(r^2(f_{13}-3f_{24})-f_{25}\bigr)=0,\qquad
V_5\bigl(r^2(f_{12}+3f_{24})-f_{26}\bigr)=0. \label{eq:det3d}
\end{align}
Equivalently, on the branches with $V_5\neq0$, these give
\[
f_{22}=f_{23}=f_5=f_6=0,
\]
together with
\[
f_{27}= -f_{13}+3f_{24},\qquad
f_{28}= -f_{12}-3f_{24},
\]
and
\[
f_{25}=r^2(f_{13}-3f_{24}),\qquad
f_{26}=r^2(f_{12}+3f_{24}).
\]
For $V_5=0$, the third-order equations impose no further restrictions.

\subsection*{Second-order determining equations}

The second-order part of the determining system is considerably longer. We
therefore give it in a reduced form, after imposing $V_4=0$ and using the
information obtained from the third-order equations. Since the third-order
equations separate the analysis into the two branches $V_5=0$ and
$V_5\neq0$, the second-order equations are also listed separately in these
two cases.

\subsubsection*{The branch $V_5=0$}

For
\[
V_5(r)=0,
\]
the third-order equations impose no additional restrictions on the coefficient
functions. In this branch, the reduced second-order determining equations are
as follows.

The equations involving $f_{10}$ and $f_{11}$ give immediately
\begin{equation}
f_{10}=f_{11}=0.
\label{eq:det2_V5zero_f10f11}
\end{equation}
After this simplification, the remaining second-order equations split naturally
into several groups. First, we have
\begin{align}
&\hbar f_{22}'+rV_1 f_{22}=0,
\qquad
\hbar f_{23}'+rV_1 f_{23}=0,
\label{eq:det2_V5zero_22_23}\\
&(r^2V_1-\hbar)f_{22}+V_1f_5=0,
\qquad
(r^2V_1-\hbar)f_{23}+V_1f_6=0,
\label{eq:det2_V5zero_5_6}\\
&f_5'=0,\qquad f_6'=0.
\label{eq:det2_V5zero_5_6_der}
\end{align}
A reduced set of the remaining equations is:
\begin{align}
&\hbar f_{19}'+2rV_1f_{19}=0,
\label{eq:det2_V5zero_f19}\\
&(\hbar-r^2V_1)f_{16}
  +V_1(f_{25}+f_{26})
  +\hbar(f_{17}+f_{27})=0,
\label{eq:det2_V5zero_alg1}\\
&f_{17}-f_{18}+f_{27}-f_{28}=0,
\label{eq:det2_V5zero_alg2}\\
&(r^2V_1-\hbar)f_{12}
  +(\hbar-r^2V_1)f_{16}
  +\hbar f_{18}
  +3(r^2V_1-\hbar)f_{24}
  +V_1f_{25}=0,
\label{eq:det2_V5zero_alg3}\\
&(r^2V_1-\hbar)f_{13}
  +(\hbar-r^2V_1)f_{16}
  +\hbar f_{17}
  -3(r^2V_1-\hbar)f_{24}
  +V_1f_{26}=0,
\label{eq:det2_V5zero_alg4}\\
&\hbar f_{16}'
  +2rV_1f_{16}
  +\hbar f_{17}'
  +rV_1(f_{17}+f_{18})
  +(3\hbar r-r^3V_1)f_{19}
  +\hbar f_{27}'
  +rV_1(f_{27}+f_{28})=0,
\label{eq:det2_V5zero_diff1}\\
&\hbar f_{16}'
  +2rV_1f_{16}
  +rV_1(f_{17}+f_{18})
  +(3\hbar r-r^3V_1)f_{19}
  +\hbar f_{18}'
  +\hbar f_{28}'
  +rV_1(f_{27}+f_{28})=0,
\label{eq:det2_V5zero_diff2}\\
&\hbar f_{13}'
  +\hbar f_{18}'
  -3\hbar f_{24}'
  +\hbar r f_{19}
  -rV_1(f_{12}+r^2f_{19}+3f_{24}+f_{28})=0,
\label{eq:det2_V5zero_diff3}\\
&\hbar f_{12}'
  +\hbar f_{17}'
  +3\hbar f_{24}'
  +\hbar r f_{19}
  -rV_1(f_{13}+r^2f_{19}-3f_{24}+f_{27})=0.
\label{eq:det2_V5zero_diff4}
\end{align}

Finally, the equations containing $f_2,f_7$, and $f_9$ reduce to
\begin{equation}
f_2'=0,\qquad f_7=0.
\label{eq:det2_V5zero_f2f7}
\end{equation}
The remaining equations involving $f_9$ may be chosen as
\begin{align}
&\hbar f_9'
+2\hbar r f_{14}
+\hbar r^2 f_{16}'
+2r^3V_1f_{16}
+r^3V_1(f_{17}+f_{18}+f_{27}+f_{28})
+2\hbar r^3 f_{19}=0,
\label{eq:det2_V5zero_f9a}\\
&\hbar f_9'
+r(\hbar-r^2V_1)(f_{12}+f_{13})
+\hbar r f_{14}
+\hbar r^2 f_{16}'
+2r^3V_1f_{16}
+\hbar r^3 f_{19}
+r^3V_1(f_{27}+f_{28})=0.
\label{eq:det2_V5zero_f9b}
\end{align}

\subsubsection*{The branch $V_5\neq0$}

For $
V_5(r)\neq0,$
we first use the relations obtained from the third-order equations,
\[
f_{22}=f_{23}=f_5=f_6=0,
\]
\[
f_{27}= -f_{13}+3f_{24},\qquad
f_{28}= -f_{12}-3f_{24},
\]
\[
f_{25}=r^2(f_{13}-3f_{24}),\qquad
f_{26}=r^2(f_{12}+3f_{24}).
\]
After substituting these relations into the second-order equations and removing
repeated equations, equations differing only by an overall sign, and common
nonzero factors, we obtain the following reduced system. This list is not meant
to be a minimal independent basis; rather, it records a reduced form of the
second-order determining equations obtained from the commutator.

In the following equations, all coefficient functions and potentials are
evaluated at $r$, and a prime denotes differentiation with respect to $r$.

First, the equations involving $f_{10},f_{11},f_{20}$, and $f_{21}$ can be
chosen as
\begin{align}
&(2r^2V_5+1)(f_{10}+f_{11})=0,
\qquad
(2r^2V_5-1)(f_{10}-f_{11})=0,
\label{eq:det2_V5nonzero_f10f11_1}\\
&f_{10}+2r^2V_5f_{11}=0,
\qquad
2r^2V_5f_{10}+f_{11}=0,
\label{eq:det2_V5nonzero_f10f11_2}\\
&f_{10}'=2rV_5f_{11},
\qquad
f_{11}'=2rV_5f_{10},
\label{eq:det2_V5nonzero_f10f11_3}\\
&\hbar f_{10}+r^2V_5(\hbar f_{11}-2f_{20})=0,
\qquad
\hbar f_{11}+r^2V_5(\hbar f_{10}-2f_{21})=0,
\label{eq:det2_V5nonzero_f20f21_1}\\
&-\hbar f_{10}'+\hbar rV_5f_{11}-2rV_5f_{20}=0,
\qquad
\hbar f_{10}'-3\hbar rV_5f_{11}-2rV_5f_{20}=0,
\label{eq:det2_V5nonzero_f20}\\
&\hbar rV_5f_{10}-\hbar f_{11}'-2rV_5f_{21}=0,
\qquad
-3\hbar rV_5f_{10}+\hbar f_{11}'-2rV_5f_{21}=0.
\label{eq:det2_V5nonzero_f21}
\end{align}
In particular, these equations imply
\[
(4r^4V_5^2-1)f_{10}=0,
\qquad
(4r^4V_5^2-1)f_{11}=0.
\]
Thus the special possibilities
\[
V_5(r)=\frac{1}{2r^2},
\qquad
V_5(r)=-\frac{1}{2r^2}
\]
are contained in the system. In the remaining case, one obtains
\[
f_{10}=f_{11}=f_{20}=f_{21}=0.
\]

The remaining second-order equations not involving $f_2,f_3,f_4,f_7$, and
$f_9$ may be represented by
\begin{align}
&\hbar\bigl(V_5'(f_{16}+f_{17}+f_{18})+f_{19}'\bigr)
+r f_{19}\bigl(\hbar rV_5'-8\hbar V_5+2V_1\bigr)=0,
\label{eq:det2_V5nonzero_main1}\\[1mm]
&-f_{12}(\hbar r^2V_5+\hbar-r^2V_1)
-3\hbar r^2V_5f_{13}
+r^2V_1f_{13}
+(3\hbar r^2V_5+\hbar-r^2V_1)f_{16}
\nonumber\\
&\hspace{1.5cm}
+2\hbar r^2V_5f_{17}
+\hbar f_{18}
+6\hbar r^2V_5f_{24}
-3\hbar f_{24}=0,
\label{eq:det2_V5nonzero_main2}\\
&-3\hbar r^2V_5f_{12}
+r^2V_1f_{12}
-f_{13}(\hbar r^2V_5+\hbar-r^2V_1)
+(3\hbar r^2V_5+\hbar-r^2V_1)f_{16}
\nonumber\\
&\hspace{1.5cm}
+\hbar f_{17}
+2\hbar r^2V_5f_{18}
-6\hbar r^2V_5f_{24}
+3\hbar f_{24}=0,
\label{eq:det2_V5nonzero_main3}\\
&-f_{12}(4\hbar r^2V_5+\hbar-2r^2V_1)
-f_{13}(4\hbar r^2V_5+\hbar-2r^2V_1)
+6\hbar r^2V_5f_{16}
+2\hbar f_{16}\nonumber\\
&\hspace{1.5cm}
-2r^2V_1f_{16}
+2\hbar r^2V_5f_{17}
+\hbar f_{17}
+2\hbar r^2V_5f_{18}
+\hbar f_{18}=0,
\label{eq:det2_V5nonzero_main4}\\
&\hbar f_{12}'
+r f_{12}(V_1-\hbar V_5)
-3\hbar rV_5f_{13}
+rV_1f_{13}
-\hbar f_{16}'
+\hbar r^2V_5'f_{16}
+6\hbar rV_5f_{16}
-2rV_1f_{16}
\nonumber\\
&\hspace{1.5cm}
+\hbar r^2V_5'f_{17}
+5\hbar rV_5f_{17}
-rV_1f_{17}
-\hbar f_{18}'
+\hbar r^2V_5'f_{18}
+3\hbar rV_5f_{18}
-rV_1f_{18}
\nonumber\\
&\hspace{1.5cm}
+f_{19}\bigl(\hbar r(r^3V_5'-5r^2V_5-3)+r^3V_1\bigr)
+3\hbar f_{24}'
+6\hbar rV_5f_{24}=0,
\label{eq:det2_V5nonzero_main5}\\
&r f_{12}(V_1-3\hbar V_5)
+\hbar f_{13}'
-\hbar rV_5f_{13}
+rV_1f_{13}
-\hbar f_{16}'
+\hbar r^2V_5'f_{16}
+6\hbar rV_5f_{16}
-2rV_1f_{16}
\nonumber\\
&\hspace{1.5cm}
-\hbar f_{17}'
+\hbar r^2V_5'f_{17}
+3\hbar rV_5f_{17}
-rV_1f_{17}
+\hbar r^2V_5'f_{18}
+5\hbar rV_5f_{18}
-rV_1f_{18}
\nonumber\\
&\hspace{1.5cm}
+f_{19}\bigl(\hbar r(r^3V_5'-5r^2V_5-3)+r^3V_1\bigr)
-3\hbar f_{24}'
-6\hbar rV_5f_{24}=0,
\label{eq:det2_V5nonzero_main6}\\
&\hbar f_{12}'
+2r f_{12}(V_1-2\hbar V_5)
+\hbar f_{13}'
-4\hbar rV_5f_{13}
+2rV_1f_{13}
-2\hbar f_{16}'
+2\hbar r^2V_5'f_{16}
\nonumber\\
&\hspace{1.5cm}
+12\hbar rV_5f_{16}
-4rV_1f_{16}
-\hbar f_{17}'
+2\hbar r^2V_5'f_{17}
+8\hbar rV_5f_{17}
-2rV_1f_{17}
\nonumber\\
&\hspace{1.5cm}
-\hbar f_{18}'
+2\hbar r^2V_5'f_{18}
+8\hbar rV_5f_{18}
-2rV_1f_{18}
\nonumber\\
&\hspace{1.5cm}
+2f_{19}\bigl(\hbar r(r^3V_5'-5r^2V_5-3)+r^3V_1\bigr)=0,
\label{eq:det2_V5nonzero_main7}\\
&(2r^2V_5-1)(f_{12}-f_{13}+f_{17}-f_{18}+6f_{24})=0,
\label{eq:det2_V5nonzero_main8}\\
&f_{12}'+2rV_5f_{12}
-f_{13}'-2rV_5f_{13}
+f_{17}'+2rV_5f_{17}
-f_{18}'-2rV_5f_{18}
+6f_{24}'+12rV_5f_{24}=0.
\label{eq:det2_V5nonzero_main9}
\end{align}

The equations containing $f_2,f_3,f_4,f_7$, and $f_9$ can be written as
follows:
\begin{align}
&\hbar f_2'
+\hbar f_9'
+2\hbar r f_7
+\hbar r^2 f_{16}'
+2\hbar r f_{14}
+2\hbar r^3 f_{19}
+r^3(\hbar V_5-V_1)(f_{12}+f_{13}) \nonumber\\
&\hspace{1.5cm}
+r^3(2V_1-6\hbar V_5)f_{16}
+r^3(V_1-5\hbar V_5)(f_{17}+f_{18})=0,
\label{eq:det2_V5nonzero_low1}\\
&\hbar f_2V_5'
+\hbar r^2f_7V_5'
-2\hbar rV_5f_7
+\hbar f_9V_5'
+\hbar f_{14}'
-\hbar f_{16}'
+r f_{12}(V_1-\hbar V_5)
\nonumber\\
&\hspace{1.5cm}
+r f_{13}(V_1-\hbar V_5)
+\hbar r^2V_5'(f_{16}+f_{17}+f_{18})
+r(6\hbar V_5-2V_1)f_{16}
\nonumber\\
&\hspace{1.5cm}
+r(5\hbar V_5-V_1)(f_{17}+f_{18})
+\hbar r^2f_{19}'
+f_{19}\bigl(\hbar r(r^3V_5'-8r^2V_5-2)+2r^3V_1\bigr)=0,
\label{eq:det2_V5nonzero_low2}\\
&r f_{12}(7\hbar r^2V_5+\hbar-2r^2V_1)
+r f_{13}(7\hbar r^2V_5+\hbar-2r^2V_1)
+\hbar r f_{14}
\nonumber\\
&\hspace{1.5cm}
+\hbar r^2 f_{16}'
-6\hbar r^3V_5f_{16}
+2r^3V_1f_{16}
-\hbar r^3V_5(f_{17}+f_{18})
+\hbar r^3f_{19}
\nonumber\\
&\hspace{1.5cm}
+\hbar f_2'
+2\hbar rV_5f_2
-2r^3V_5(f_3+f_4)
+\hbar r f_7
+\hbar f_9'
+2\hbar rV_5f_9=0,
\label{eq:det2_V5nonzero_low3}\\
&-\hbar f_{12}'
+r f_{12}(2V_1-3\hbar V_5)
-\hbar f_{13}'
-3\hbar rV_5f_{13}
+2rV_1f_{13}
+2\hbar f_{14}'
+2\hbar r^2V_5'f_{14}
-2\hbar rV_5f_{14}
\nonumber\\
&\hspace{1.5cm}
-\hbar f_{16}'
+2\hbar r^2V_5'f_{16}
+6\hbar rV_5f_{16}
-2rV_1f_{16}
+2\hbar r^2V_5'(f_{17}+f_{18})
+\hbar rV_5(f_{17}+f_{18})
\nonumber\\
&\hspace{1.5cm}
+2\hbar r^2f_{19}'
+f_{19}\bigl(\hbar r(2r^3V_5'-16r^2V_5-1)+4r^3V_1\bigr)
+2\hbar f_2V_5'
\nonumber\\
&\hspace{1.5cm}
+2rV_5(f_3+f_4)
+2\hbar f_7'
+2\hbar r^2V_5'f_7
-2\hbar rV_5f_7
+2\hbar f_9V_5'=0,
\label{eq:det2_V5nonzero_low4}\\
&-f_{12}(3\hbar r^2V_5+\hbar-r^2V_1)
-4\hbar r^2V_5f_{13}
+r^2V_1f_{13}
+(3\hbar r^2V_5+\hbar-r^2V_1)f_{16}
\nonumber\\
&\hspace{1.5cm}
+\hbar r^2V_5f_{17}
+\hbar f_{18}
-\hbar V_5f_2
+3\hbar r^2V_5f_{24}
-3\hbar f_{24}
+2r^2V_5f_4
-\hbar V_5f_9=0,
\label{eq:det2_V5nonzero_low5}\\
&-4\hbar r^2V_5f_{12}
+r^2V_1f_{12}
-f_{13}(3\hbar r^2V_5+\hbar-r^2V_1)
+(3\hbar r^2V_5+\hbar-r^2V_1)f_{16}
\nonumber\\
&\hspace{1.5cm}
+\hbar f_{17}
+\hbar r^2V_5f_{18}
-\hbar V_5f_2
-3\hbar r^2V_5f_{24}
+3\hbar f_{24}
+2r^2V_5f_3
-\hbar V_5f_9=0,
\label{eq:det2_V5nonzero_low6}\\
&-\hbar f_{13}'
-\hbar rV_5f_{13}
+\hbar r^2V_5'f_{14}
-\hbar rV_5f_{14}
+\hbar r^2V_5'f_{16}
+\hbar rV_5f_{16}
+\hbar r^2V_5'f_{17}
-\hbar f_{18}'
\nonumber\\
&\hspace{1.5cm}
+\hbar r^2V_5'f_{18}
-\hbar rV_5f_{18}
+f_{19}\bigl(\hbar r(r^3V_5'-4r^2V_5-1)+r^3V_1\bigr)
+\hbar f_2V_5'
+3\hbar f_{24}'
\nonumber\\
&\hspace{1.5cm}
+3\hbar rV_5f_{24}
+2rV_5f_4
+\hbar r^2V_5'f_7
-\hbar rV_5f_7
+\hbar f_9V_5'=0,
\label{eq:det2_V5nonzero_low7}\\
&-\hbar f_{12}'
-\hbar rV_5f_{12}
+\hbar r^2V_5'f_{14}
-\hbar rV_5f_{14}
+\hbar r^2V_5'f_{16}
+\hbar rV_5f_{16}
-\hbar f_{17}'
+\hbar r^2V_5'f_{17}
\nonumber\\
&\hspace{1.5cm}
-\hbar rV_5f_{17}
+\hbar r^2V_5'f_{18}
+f_{19}\bigl(\hbar r(r^3V_5'-4r^2V_5-1)+r^3V_1\bigr)
+\hbar f_2V_5'
-3\hbar f_{24}'
\nonumber\\
&\hspace{1.5cm}
-3\hbar rV_5f_{24}
+2rV_5f_3
+\hbar r^2V_5'f_7
-\hbar rV_5f_7
+\hbar f_9V_5'=0.
\label{eq:det2_V5nonzero_low8}
\end{align}

\subsection*{First-order determining equations}

We now list the determining equations obtained from the first-order part of
the commutator.
The following list is obtained after omitting equations which differ only by an
overall sign or are immediate repetitions of previously listed equations. The
system is not reduced to a minimal independent basis.

A convenient reduced list of first-order determining equations is:
\begin{align}
&2\hbar f_1
+2\hbar^2r^2V_5 f_{10}
+\hbar r^2V_1f_{10}
+2\hbar^2r^2V_5f_{11}
+\hbar r^2V_1f_{11}
+2\hbar r^2f_{15}
\nonumber\\
&\quad
-8\hbar r^2V_5f_{20}
+2r^2V_1f_{20}
-8\hbar r^2V_5f_{21}
+2r^2V_1f_{21}
-\hbar^2r^2V_5f_{22}
-2\hbar r^2V_1f_{22}
\nonumber\\
&\quad
-\hbar^2r^2V_5f_{23}
-2\hbar r^2V_1f_{23}
+\hbar^2rV_5f_5'
+\hbar^2rf_5V_5'
+\hbar^2V_5f_5
-\hbar rf_5V_1'
-3\hbar V_1f_5
\nonumber\\
&\quad
+\hbar^2rV_5f_6'
+\hbar^2rf_6V_5'
+\hbar^2V_5f_6
-\hbar rf_6V_1'
-3\hbar V_1f_6
+2\hbar f_8=0,
\label{eq:det1_1}\\
&-2\hbar f_1'
+rf_{10}(2\hbar^2V_5+\hbar V_1+4r^2V_3)
+2\hbar^2rV_5f_{11}
+\hbar rV_1f_{11}
+4r^3V_3f_{11}
\nonumber\\
&\quad
-2\hbar r^2f_{15}'
+2rf_{15}(8\hbar r^2V_5+\hbar-2r^2V_1)
-8\hbar rV_5f_{20}
+2rV_1f_{20}
\nonumber\\
&\quad
-8\hbar rV_5f_{21}
+2rV_1f_{21}
-\hbar^2rV_5f_{22}
-2\hbar rV_1f_{22}
-\hbar^2rV_5f_{23}
-2\hbar rV_1f_{23}
\nonumber\\
&\quad
+\hbar^2V_5f_5'
+\hbar^2f_5V_5'
-\hbar f_5V_1'
+\hbar^2V_5f_6'
+\hbar^2f_6V_5'
-\hbar f_6V_1'
-2\hbar f_8'=0,
\label{eq:det1_2}\\
&2\hbar f_1'
-4\hbar rV_5f_1
+rf_{10}(3\hbar^2V_5-2\hbar V_1-4r^2V_3)
+rf_{11}(3\hbar^2V_5-2\hbar V_1-4r^2V_3)
\nonumber\\
&\quad
+2\hbar r^2f_{15}'
-16\hbar r^3V_5f_{15}
+4r^3V_1f_{15}
+2\hbar rV_5f_{20}
+2\hbar rV_5f_{21}
\nonumber\\
&\quad
-2\hbar^2rV_5f_{22}
+\hbar rV_1f_{22}
-2\hbar^2rV_5f_{23}
+\hbar rV_1f_{23}
+\hbar^2V_5f_5'
+2\hbar f_5V_1'
\nonumber\\
&\quad
+\hbar^2V_5f_6'
+2\hbar f_6V_1'
+2\hbar f_8'
-4\hbar rV_5f_8=0,
\label{eq:det1_3}\\
&2rV_3f_{10}
+2rV_3f_{11}
-\hbar f_{15}'
-2r(V_1-4\hbar V_5)f_{15}=0,
\label{eq:det1_4}\\
&\hbar rf_{12}(3V_1-14\hbar V_5)
+\hbar rf_{13}(3V_1-14\hbar V_5)
+2\hbar r^2f_{14}V_1'
-2\hbar rV_1f_{14}
\nonumber\\
&\quad
+2\hbar^2r^2V_5'f_{16}
+2\hbar^2rV_5f_{16}
+2\hbar^2r^2V_5'f_{17}
-3\hbar^2rV_5f_{17}
+4r^3V_3f_{17}
\nonumber\\
&\quad
+2\hbar^2r^2V_5'f_{18}
-3\hbar^2rV_5f_{18}
+4r^3V_3f_{18}
+2\hbar^2r^4V_5'f_{19}
-2\hbar^2r^3V_5f_{19}
\nonumber\\
&\quad
+2\hbar f_2V_1'
-\hbar^2V_5f_{25}'
+8rV_3f_{25}
-\hbar^2V_5f_{26}'
+8rV_3f_{26}
-\hbar^2r^2V_5f_{27}'
\nonumber\\
&\quad
-11\hbar^2rV_5f_{27}
+\hbar rV_1f_{27}
+4r^3V_3f_{27}
-\hbar^2r^2V_5f_{28}'
-11\hbar^2rV_5f_{28}
\nonumber\\
&\quad
+\hbar rV_1f_{28}
+4r^3V_3f_{28}
-2\hbar f_3'
+6\hbar rV_5f_3
-2\hbar f_4'
+6\hbar rV_5f_4
\nonumber\\
&\quad
+2\hbar r^2f_7V_1'
-2\hbar rV_1f_7
+2\hbar f_9V_1'=0,
\label{eq:det1_5}\\
&-\hbar r^3f_{10}''
-7\hbar r^2f_{10}'
+11\hbar r^3V_5f_{10}
-\hbar r^3f_{11}''
-7\hbar r^2f_{11}'
+11\hbar r^3V_5f_{11}
\nonumber\\
&\quad
+2r^3V_5f_{20}
+2r^3V_5f_{21}
+\hbar r^2f_{22}'
+r^3(V_1-4\hbar V_5)f_{22}
+\hbar r^2f_{23}'
+r^3(V_1-4\hbar V_5)f_{23}
\nonumber\\
&\quad
-\hbar rf_5''
+3\hbar r^2V_5f_5'
+\hbar f_5'
-2\hbar r^2V_5'f_5
-\hbar rf_6''
+3\hbar r^2V_5f_6'
+\hbar f_6'
-2\hbar r^2V_5'f_6=0,
\label{eq:det1_6}\\
&\hbar rf_{16}''
-f_{16}'(8\hbar r^2V_5+\hbar-2r^2V_1)
+2\hbar r^2V_5'f_{16}
+2r^2V_1'f_{16}
\nonumber\\
&\quad
+\hbar rf_{17}''
-8\hbar r^2V_5f_{17}'
-\hbar f_{17}'
+2r^2V_1f_{17}'
+2\hbar r^2V_5'f_{17}
+2r^2V_1'f_{17}
\nonumber\\
&\quad
+\hbar rf_{18}''
-8\hbar r^2V_5f_{18}'
-\hbar f_{18}'
+2r^2V_1f_{18}'
+2\hbar r^2V_5'f_{18}
+2r^2V_1'f_{18}
+2\hbar r^3f_{19}''
\nonumber\\
&\quad
-8\hbar r^4V_5f_{19}'
+17\hbar r^2f_{19}'
+2r^4V_1f_{19}'
+2r^3f_{19}\bigl(\hbar rV_5'-48\hbar V_5+rV_1'+12V_1\bigr)=0,
\label{eq:det1_7}\\
&\hbar\Bigl(
7\hbar r^2V_5f_{12}
+7\hbar r^2V_5f_{13}
-2r^2(V_1-3\hbar V_5)f_{16}
+r^2(V_1-4\hbar V_5)f_{17}
\nonumber\\
&\quad
-4\hbar r^2V_5f_{18}
+r^2V_1f_{18}
-2V_1f_2
+3\hbar rV_5f_{25}'
-\hbar rV_5'f_{25}
-5\hbar V_5f_{25}
+rV_1'f_{25}
+V_1f_{25}
\nonumber\\
&\quad
+3\hbar rV_5f_{26}'
-\hbar rV_5'f_{26}
-5\hbar V_5f_{26}
+rV_1'f_{26}
+V_1f_{26}
+3\hbar r^3V_5f_{27}'
-\hbar r^3V_5'f_{27}
\nonumber\\
&\quad
+18\hbar r^2V_5f_{27}
+r^3V_1'f_{27}
+3\hbar r^3V_5f_{28}'
-\hbar r^3V_5'f_{28}
+18\hbar r^2V_5f_{28}
+r^3V_1'f_{28}
-2V_1f_9
\Bigr)
\nonumber\\
&\quad
-2(\hbar-r^2V_1)f_3
-2(\hbar-r^2V_1)f_4=0,
\label{eq:det1_8}\\
&\hbar rf_{10}(r^2V_5-1)
+\hbar rf_{11}(r^2V_5-1)
+6r^3V_5f_{20}
+6r^3V_5f_{21}
+\hbar r^3f_{22}''
+6\hbar r^2f_{22}'
\nonumber\\
&\quad
+4\hbar r^3V_5f_{22}
+\hbar rf_{22}
+3r^3V_1f_{22}
+\hbar r^3f_{23}''
+6\hbar r^2f_{23}'
+4\hbar r^3V_5f_{23}
+\hbar rf_{23}
+3r^3V_1f_{23}
\nonumber\\
&\quad
+\hbar rf_5''
+3\hbar f_5'
-r^2V_1f_5'
+\hbar r^2V_5'f_5
+3\hbar rV_5f_5
-r^2V_1'f_5
-rV_1f_5
\nonumber\\
&\quad
+\hbar rf_6''
+3\hbar f_6'
-r^2V_1f_6'
+\hbar r^2V_5'f_6
+3\hbar rV_5f_6
-r^2V_1'f_6
-rV_1f_6=0,
\label{eq:det1_9}\\
&\hbar rf_{12}(\hbar V_5+3V_1)
+\hbar rf_{13}(\hbar V_5+3V_1)
+2\hbar r^2f_{14}V_1'
+2\hbar^2r^2V_5'f_{16}
+10\hbar^2rV_5f_{16}
-2\hbar rV_1f_{16}
\nonumber\\
&\quad
+2\hbar^2r^2V_5'f_{17}
-5\hbar^2rV_5f_{17}
+\hbar rV_1f_{17}
+4r^3V_3f_{17}
+2\hbar^2r^2V_5'f_{18}
-5\hbar^2rV_5f_{18}
+\hbar rV_1f_{18}
\nonumber\\
&\quad
+4r^3V_3f_{18}
+2\hbar^2r^4V_5'f_{19}
+2\hbar f_2V_1'
+4\hbar^2V_5f_{25}'
+\hbar^2V_5'f_{25}
+\hbar V_1'f_{25}
+4rV_3f_{25}
+4\hbar^2V_5f_{26}'
\nonumber\\
&\quad
+\hbar^2V_5'f_{26}
+\hbar V_1'f_{26}
+4rV_3f_{26}
+4\hbar^2r^2V_5f_{27}'
+\hbar^2r^2V_5'f_{27}
+24\hbar^2rV_5f_{27}
+\hbar r^2V_1'f_{27}\nonumber\\
&\quad
+4\hbar^2r^2V_5f_{28}'
+\hbar^2r^2V_5'f_{28}
+24\hbar^2rV_5f_{28}
+\hbar r^2V_1'f_{28}
+6\hbar rV_5f_3
+2rV_1f_3 \nonumber\\
&\quad
+6\hbar rV_5f_4
+2rV_1f_4
+2\hbar r^2f_7V_1'
+2\hbar f_9V_1'=0,
\label{eq:det1_10}\\
&\hbar rf_{10}''
+6\hbar f_{10}'
-4\hbar rV_5f_{10}
+\hbar rf_{11}''
+6\hbar f_{11}'
-4\hbar rV_5f_{11}
+8rV_5f_{20}
+8rV_5f_{21}
\nonumber\\
&\quad
+\hbar rf_{22}''
+6\hbar f_{22}'
+4r(2\hbar V_5+V_1)f_{22}
+\hbar rf_{23}''
+6\hbar f_{23}'
+4r(2\hbar V_5+V_1)f_{23}
\nonumber\\
&\quad
-3\hbar V_5f_5'
-V_1f_5'
+\hbar V_5'f_5
-V_1'f_5
-3\hbar V_5f_6'
-V_1f_6'
+\hbar V_5'f_6
-V_1'f_6=0,
\label{eq:det1_11}\\
&-\hbar^2r^3f_{10}''
-7\hbar^2r^2f_{10}'
+r^3(2\hbar^2V_5-\hbar V_1+4r^2V_3+4V_2)f_{10}
+10\hbar^2r^3V_5f_{11}
\nonumber\\
&\quad
+\hbar r^3V_1f_{11}
-4r^3V_2f_{11}
+2f_{15}\bigl(\hbar r^3(5r^2V_5+2)-r^5V_1\bigr)
-4\hbar r^3V_5f_{20}
+2r^3V_1f_{20}
\nonumber\\
&\quad
+2\hbar r^2f_{21}'
+2r^3V_1f_{21}
+\hbar^2r^2f_{22}'
-\hbar^2r^3V_5f_{22}
-3\hbar^2r^3V_5f_{23}
-\hbar^2rf_5''
+\hbar^2r^2V_5f_5'
\nonumber\\
&\quad
+\hbar^2f_5'
+2\hbar^2r^2V_5'f_5
+3\hbar^2r^2V_5f_6'
-2\hbar^2r^2V_5'f_6=0,
\label{eq:det1_12}\\
&\hbar^2rf_{10}'
-\hbar^2f_{10}
-6\hbar^2r^2V_5f_{11}
-4\hbar r^2V_5f_{20}
-2\hbar rf_{21}'
-2\hbar f_{21}
-\hbar^2rf_{22}'
+\hbar^2f_{22}
\nonumber\\
&\quad
-2\hbar r^2V_1f_{22}
-8\hbar^2r^2V_5f_{23}
+2\hbar^2f_5''
+\frac{2\hbar^2}{r}f_5'
+\hbar^2V_5f_5
-\hbar V_1f_5
-4r^2V_3f_5
-4V_2f_5
\nonumber\\
&\quad
+2\hbar^2rV_5f_6'
+4\hbar^2rV_5'f_6
+5\hbar^2V_5f_6
-\hbar V_1f_6
+4V_2f_6=0,
\label{eq:det1_14}\\
&2\hbar rV_5f_1
+\hbar^2rf_{10}''
+6\hbar^2f_{10}'
-r(5\hbar^2V_5-2\hbar V_1+4r^2V_3+4V_2)f_{10}
\nonumber\\
&\quad
-4\hbar^2rV_5f_{11}
-\hbar rV_1f_{11}
+4rV_2f_{11}
+2\hbar r^2f_{15}'
+2rf_{15}(-4\hbar r^2V_5+\hbar+r^2V_1)
\nonumber\\
&\quad
+14\hbar rV_5f_{20}
-2rV_1f_{20}
+\hbar^2rf_{22}''
+6\hbar^2f_{22}'
-13\hbar^2rV_5f_{22}
+6\hbar rV_1f_{22}
-4rV_2f_{22}
\nonumber\\
&\quad
+16\hbar^2rV_5f_{23}
-\hbar rV_1f_{23}
+4rV_2f_{23}
+2\hbar^2V_5f_5'
-\hbar V_1f_5'
+\hbar^2V_5'f_5
-\hbar V_1'f_5
\nonumber\\
&\quad
-5\hbar^2V_5f_6'
+\hbar^2V_5'f_6
-\hbar V_1'f_6
-4rV_3f_6
+2\hbar rV_5f_8=0,
\label{eq:det1_15}\\
&-2\hbar f_1
+2\hbar^2r^2V_5f_{10}
-\hbar r^2V_1f_{10}
+2\hbar^2r^2V_5f_{11}
-\hbar r^2V_1f_{11}
+2\hbar r^2f_{15}
\nonumber\\
&\quad
-4\hbar r^2V_5f_{20}
+2r^2V_1f_{20}
-4\hbar r^2V_5f_{21}
+2r^2V_1f_{21}
-\hbar^2r^2V_5f_{22}
-\hbar^2r^2V_5f_{23}
\nonumber\\
&\quad
+\hbar^2rV_5f_5'
+\hbar^2rV_5'f_5
+\hbar^2V_5f_5
+\hbar rV_1'f_5
+\hbar V_1f_5
+\hbar^2rV_5f_6'
+\hbar^2rV_5'f_6
+\hbar^2V_5f_6
\nonumber\\
&\quad
+\hbar rV_1'f_6
+\hbar V_1f_6
-8\hbar r^2V_5f_8
+2\hbar f_8=0,
\label{eq:det1_16}\\
&2\hbar f_1'
+rf_{10}(2\hbar^2V_5-\hbar V_1+4r^2V_3)
+2\hbar^2rV_5f_{11}
-\hbar rV_1f_{11}
+4r^3V_3f_{11}
-2\hbar r^2f_{15}'-4\hbar rV_5f_{20}
\nonumber\\
&\quad
+2rf_{15}(8\hbar r^2V_5+\hbar-2r^2V_1)
+2rV_1f_{20}
-4\hbar rV_5f_{21}
+2rV_1f_{21}
-\hbar^2rV_5f_{22}
-\hbar^2rV_5f_{23}
\nonumber\\
&\quad
+\hbar^2V_5f_5'
+\hbar^2V_5'f_5
+\hbar V_1'f_5
+\hbar^2V_5f_6'
+\hbar^2V_5'f_6
+\hbar V_1'f_6
-2\hbar f_8'
-8\hbar rV_5f_8=0,
\label{eq:det1_17}\\
&\hbar rf_{12}(2\hbar V_5+V_1)
-\hbar rf_{13}(2\hbar V_5+V_1)
-3\hbar^2rV_5f_{17}
+2\hbar rV_1f_{17}
-4r^3V_3f_{17}
-8rV_2f_{17}
\nonumber\\
&\quad
+3\hbar^2rV_5f_{18}
-2\hbar rV_1f_{18}
+4r^3V_3f_{18}
+8rV_2f_{18}
+12\hbar^2rV_5f_{24}
+6\hbar rV_1f_{24}
\nonumber\\
&\quad
-\hbar^2V_5f_{25}'
-2\hbar^2V_5'f_{25}
+\hbar^2V_5f_{26}'
+2\hbar^2V_5'f_{26}
-\hbar^2r^2V_5f_{27}'
-2\hbar^2r^2V_5'f_{27}
\nonumber\\
&\quad
-7\hbar^2rV_5f_{27}
+\hbar rV_1f_{27}
-4r^3V_3f_{27}
-8rV_2f_{27}
+\hbar^2r^2V_5f_{28}'
+2\hbar^2r^2V_5'f_{28}
\nonumber\\
&\quad
+7\hbar^2rV_5f_{28}
-\hbar rV_1f_{28}
+4r^3V_3f_{28}
+8rV_2f_{28}
-2\hbar f_3'
-6\hbar rV_5f_3
+2\hbar f_4'
+6\hbar rV_5f_4=0,
\label{eq:det1_19}\\
&\hbar r^3f_{10}''
+7\hbar r^2f_{10}'
+11\hbar r^3V_5f_{10}
-\hbar r^3f_{11}''
-7\hbar r^2f_{11}'
-11\hbar r^3V_5f_{11}
\nonumber\\
&\quad
-2r^3V_5f_{20}
+2r^3V_5f_{21}
-\hbar r^2f_{22}'
-r^3(4\hbar V_5+V_1)f_{22}
+\hbar r^2f_{23}'
+r^3(4\hbar V_5+V_1)f_{23}
\nonumber\\
&\quad
+\hbar rf_5''
+3\hbar r^2V_5f_5'
-\hbar f_5'
-2\hbar r^2V_5'f_5
-\hbar rf_6''
-3\hbar r^2V_5f_6'
+\hbar f_6'
+2\hbar r^2V_5'f_6=0.
\label{eq:det1_20}
\end{align}

\subsection*{Zeroth-order determining equations}

Finally, the zeroth-order part of the commutator gives the following equations.
We omit equations which differ only by an overall sign or are immediate
repetitions of the displayed ones. The system is not reduced to a minimal
independent basis.

A convenient reduced list of zeroth-order determining equations is:
\begin{align}
&-2\hbar r f_1''
-8\hbar f_1'
+4\hbar rV_5f_1
+f_{10}(-\hbar^2 rV_5+3\hbar rV_1+12r^3V_3)
-\hbar^2 rV_5f_{11}
+3\hbar rV_1f_{11}
\nonumber\\
&\quad
+12r^3V_3f_{11}
-2\hbar r^3f_{15}''
-16\hbar r^2f_{15}'
-4rf_{15}(-12\hbar r^2V_5+\hbar+3r^2V_1)
-10\hbar rV_5f_{20}
\nonumber\\
&\quad
+2rV_1f_{20}
-10\hbar rV_5f_{21}
+2rV_1f_{21}
+\hbar^2 rV_5f_{22}
-3\hbar rV_1f_{22}
+\hbar^2 rV_5f_{23}
-3\hbar rV_1f_{23}
\nonumber\\
&\quad
-\hbar^2V_5f_5'
+3\hbar V_1f_5'
+4r^2V_3f_5'
+4r^2V_3'f_5
+4rV_3f_5
-\hbar^2V_5f_6'
+3\hbar V_1f_6'
+4r^2V_3f_6'
\nonumber\\
&\quad
+4r^2V_3'f_6
+4rV_3f_6
-2\hbar r f_8''
-8\hbar f_8'
+4\hbar rV_5f_8=0,
\label{eq:det0_1}\\
&8rV_3(f_{10}+f_{11})
-\hbar r f_{15}''
-8\hbar f_{15}'
-8r(V_1-4\hbar V_5)f_{15}
+2V_3(f_5'+f_6')
+2V_3'(f_5+f_6)=0,
\label{eq:det0_2}\\
&12\hbar r^3V_3(f_{12}+f_{13})
-4\hbar r^4V_3'f_{14}
-8\hbar r^3V_3f_{14}
-\hbar^3r^2f_{16}^{(3)}
-6\hbar^3r f_{16}''
+32\hbar^3r^2V_5f_{16}'
+6\hbar^3f_{16}'
\nonumber\\
&\quad
-8\hbar^2r^2V_1f_{16}'
-4\hbar r^2(V_0'+V_2')f_{16}
-\hbar^3r^2f_{17}^{(3)}
-6\hbar^3r f_{17}''
+32\hbar^3r^2V_5f_{17}'
+6\hbar^3f_{17}'
\nonumber\\
&\quad
-8\hbar^2r^2V_1f_{17}'
+4\hbar r^3V_3f_{17}
-4\hbar r^2(V_0'+V_2')f_{17}
-\hbar^3r^2f_{18}^{(3)}
-6\hbar^3r f_{18}''
+32\hbar^3r^2V_5f_{18}'
\nonumber\\
&\quad
+6\hbar^3f_{18}'
-8\hbar^2r^2V_1f_{18}'
+4\hbar r^3V_3f_{18}
-4\hbar r^2(V_0'+V_2')f_{18}
-\hbar^3r^4f_{19}^{(3)}
-16\hbar^3r^3f_{19}''
\nonumber\\
&\quad
+32\hbar^3r^4V_5f_{19}'
-56\hbar^3r^2f_{19}'
-8\hbar^2r^4V_1f_{19}'
+4\hbar r^3f_{19}\bigl(48\hbar^2V_5-12\hbar V_1-r(V_0'+V_2')\bigr)
\nonumber\\
&\quad
-4\hbar r^2V_3'f_2
-8r^3V_3(f_3+f_4)
-4\hbar r^4V_3'f_7
-8\hbar r^3V_3f_7
-4\hbar r^2V_3'f_9=0,
\label{eq:det0_3}\\
&\hbar r(5\hbar V_5-3V_1)(f_{12}+f_{13})
+2\hbar r^2V_1f_{14}'
+8\hbar rV_1f_{14}
+2\hbar^2r^2V_5f_{16}'
+10\hbar^2rV_5f_{16}
-2\hbar rV_1f_{16}
\nonumber\\
&\quad
+2\hbar^2r^2V_5f_{17}'
+5\hbar^2rV_5f_{17}
+\hbar rV_1f_{17}
-4r^3V_3f_{17}
+2\hbar^2r^2V_5f_{18}'
+5\hbar^2rV_5f_{18}
+\hbar rV_1f_{18}
\nonumber\\
&\quad
-4r^3V_3f_{18}
+2\hbar^2r^4V_5f_{19}'
+12\hbar^2r^3V_5f_{19}
+2\hbar V_1f_2'
+5\hbar^2V_5f_{25}'
-\hbar V_1f_{25}'
-4r^2V_3f_{25}'
\nonumber\\
&\quad
-4r^2V_3'f_{25}
-20rV_3f_{25}
+5\hbar^2V_5f_{26}'
-\hbar V_1f_{26}'
-4r^2V_3f_{26}'
-4r^2V_3'f_{26}
-20rV_3f_{26}
\nonumber\\
&\quad
+5\hbar^2r^2V_5f_{27}'
-\hbar r^2V_1f_{27}'
-4r^4V_3f_{27}'
+20\hbar^2rV_5f_{27}
-4\hbar rV_1f_{27}
-4r^4V_3'f_{27}
-24r^3V_3f_{27}
\nonumber\\
&\quad
+5\hbar^2r^2V_5f_{28}'
-\hbar r^2V_1f_{28}'
-4r^4V_3f_{28}'
+20\hbar^2rV_5f_{28}
-4\hbar rV_1f_{28}
-4r^4V_3'f_{28}
-24r^3V_3f_{28}
\nonumber\\
&\quad
+2\hbar r f_3''
+8\hbar f_3'
-6\hbar rV_5f_3
+2rV_1f_3
+2\hbar r f_4''
+8\hbar f_4'
-6\hbar rV_5f_4
+2rV_1f_4
\nonumber\\
&\quad
+2\hbar r^2V_1f_7'
+8\hbar rV_1f_7
+2\hbar V_1f_9'=0,
\label{eq:det0_4}\\
&4\hbar rV_1f_1
-\hbar^3r f_{10}''
-4\hbar^3f_{10}'
+\hbar^2r(3\hbar V_5-V_1)f_{10}
-\hbar^3r f_{11}''
-4\hbar^3f_{11}'
\nonumber\\
&\quad
+3\hbar^3rV_5f_{11}
-\hbar^2rV_1f_{11}
+4\hbar^2r^3V_5f_{15}
+10\hbar^2rV_5f_{20}
-2\hbar rV_1f_{20}
-8r^3V_3f_{20}
\nonumber\\
&\quad
+10\hbar^2rV_5f_{21}
-2\hbar rV_1f_{21}
-8r^3V_3f_{21}
+\hbar^3r f_{22}''
+4\hbar^3f_{22}'
\nonumber\\
&\quad
-3\hbar^3rV_5f_{22}
+\hbar^2rV_1f_{22}
+4\hbar r^3V_3f_{22}
+\hbar^3r f_{23}''
+4\hbar^3f_{23}'
\nonumber\\
&\quad
-3\hbar^3rV_5f_{23}
+\hbar^2rV_1f_{23}
+4\hbar r^3V_3f_{23}
-\hbar^3f_5^{(3)}
-\frac{2\hbar^3}{r}f_5''
+\frac{2\hbar^3}{r^2}f_5'
\nonumber\\
&\quad
+3\hbar^3V_5f_5'
-\hbar^2V_1f_5'
-4\hbar(V_0'+V_2')f_5
+4\hbar rV_3f_5
-\hbar^3f_6^{(3)}
-\frac{2\hbar^3}{r}f_6''
+\frac{2\hbar^3}{r^2}f_6'
\nonumber\\
&\quad
+3\hbar^3V_5f_6'
-\hbar^2V_1f_6'
-4\hbar(V_0'+V_2')f_6
+4\hbar rV_3f_6
+4\hbar rV_1f_8=0,
\label{eq:det0_5}\\
&f_{19}V_3'=0,
\label{eq:det0_6}\\
&-2\hbar r(\hbar V_5+V_1)f_1
+\hbar^3r f_{10}''
+4\hbar^3f_{10}'
+\hbar^3rV_5f_{10}
+\hbar^2rV_1f_{10}
-4\hbar r^3V_3f_{10}
-4\hbar rV_2f_{10}
\nonumber\\
&\quad
-\hbar^3rV_5f_{11}
-\hbar^2rV_1f_{11}
+4\hbar rV_2f_{11}
-2\hbar r f_{15}\bigl(\hbar(7r^2V_5+2)-r^2V_1\bigr)
\nonumber\\
&\quad
-8rV_2f_{20}
-2\hbar^2r f_{21}''
-8\hbar^2f_{21}'
-4\hbar^2rV_5f_{21}
-4\hbar rV_1f_{21}
+8r^3V_3f_{21}
+8rV_2f_{21}
\nonumber\\
&\quad
-\hbar^3r f_{22}''
-4\hbar^3f_{22}'
-\hbar^3rV_5f_{22}
-\hbar^2rV_1f_{22}
+4\hbar rV_2f_{22}
+\hbar^3rV_5f_{23}
+\hbar^2rV_1f_{23}
\nonumber\\
&\quad
-4\hbar r^3V_3f_{23}
-4\hbar rV_2f_{23}
+\hbar^3f_5^{(3)}
+\frac{2\hbar^3}{r}f_5''
-\frac{2\hbar^3}{r^2}f_5'
+\hbar^3V_5f_5'
+\hbar^2V_1f_5'
\nonumber\\
&\quad
-4\hbar r^2V_3f_5'
-4\hbar V_2f_5'
-4\hbar r^2V_3'f_5
+4\hbar V_0'f_5
-4\hbar V_2'f_5
-12\hbar rV_3f_5
\nonumber\\
&\quad
-\hbar^3V_5f_6'
-\hbar^2V_1f_6'
+4\hbar V_2f_6'
+8\hbar V_2'f_6
-4\hbar rV_3f_6
-2\hbar^2rV_5f_8
-2\hbar rV_1f_8=0,
\label{eq:det0_7}\\
&r(\hbar^2V_5-\hbar V_1+8V_2)f_{12}
+r(-\hbar^2V_5+\hbar V_1-8V_2)f_{13}
-\hbar^2rV_5f_{17}
+\hbar rV_1f_{17}
-4r^3V_3f_{17}
\nonumber\\
&\quad
-8rV_2f_{17}
+\hbar^2rV_5f_{18}
-\hbar rV_1f_{18}
+4r^3V_3f_{18}
+8rV_2f_{18}
+6\hbar^2rV_5f_{24}
-6\hbar rV_1f_{24}
\nonumber\\
&\quad
+48rV_2f_{24}
-\hbar^2V_5f_{25}'
+\hbar V_1f_{25}'
-4r^2V_3f_{25}'
-8V_2f_{25}'
-4r^2V_3'f_{25}
-8V_2'f_{25}
-12rV_3f_{25}
\nonumber\\
&\quad
+\hbar^2V_5f_{26}'
-\hbar V_1f_{26}'
+4r^2V_3f_{26}'
+8V_2f_{26}'
+4r^2V_3'f_{26}
+8V_2'f_{26}
+12rV_3f_{26}
-\hbar^2r^2V_5f_{27}'
\nonumber\\
&\quad
+\hbar r^2V_1f_{27}'
-4r^4V_3f_{27}'
-8r^2V_2f_{27}'
-4\hbar^2rV_5f_{27}
+4\hbar rV_1f_{27}
-4r^4V_3'f_{27}
-24r^3V_3f_{27}
\nonumber\\
&\quad
-8r^2V_2'f_{27}
-32rV_2f_{27}
+\hbar^2r^2V_5f_{28}'
-\hbar r^2V_1f_{28}'
+4r^4V_3f_{28}'
+8r^2V_2f_{28}'
+4\hbar^2rV_5f_{28}
\nonumber\\
&\quad
-4\hbar rV_1f_{28}
+4r^4V_3'f_{28}
+24r^3V_3f_{28}
+8r^2V_2'f_{28}
+32rV_2f_{28}
-2\hbar r f_3''
-8\hbar f_3'
\nonumber\\
&\quad
-6\hbar rV_5f_3
-2rV_1f_3
+2\hbar r f_4''
+8\hbar f_4'
+6\hbar rV_5f_4
+2rV_1f_4=0,
\label{eq:det0_8}\\
&\hbar^3r f_{10}''
+4\hbar^3f_{10}'
+\hbar^2r(3\hbar V_5+V_1)f_{10}
-\hbar^3r f_{11}''
-4\hbar^3f_{11}'
-\hbar^2r(3\hbar V_5+V_1)f_{11}
\nonumber\\
&\quad
-2\hbar^2rV_5f_{20}
+2\hbar rV_1f_{20}
-8r^3V_3f_{20}
-16rV_2f_{20}
+2\hbar^2rV_5f_{21}
-2\hbar rV_1f_{21}
+8r^3V_3f_{21}
+16rV_2f_{21}
\nonumber\\
&\quad
-\hbar^3r f_{22}''
-4\hbar^3f_{22}'
-3\hbar^3rV_5f_{22}
-\hbar^2rV_1f_{22}
+4\hbar r^3V_3f_{22}
+\hbar^3r f_{23}''
+4\hbar^3f_{23}'
\nonumber\\
&\quad
+3\hbar^3rV_5f_{23}
+\hbar^2rV_1f_{23}
-4\hbar r^3V_3f_{23}
+\hbar^3f_5^{(3)}
+\frac{2\hbar^3}{r}f_5''
-\frac{2\hbar^3}{r^2}f_5'
\nonumber\\
&\quad
+3\hbar^3V_5f_5'
+\hbar^2V_1f_5'
+4\hbar V_0'f_5
-4\hbar V_2'f_5
+4\hbar rV_3f_5
-\hbar^3f_6^{(3)}
-\frac{2\hbar^3}{r}f_6''
+\frac{2\hbar^3}{r^2}f_6'
\nonumber\\
&\quad
-3\hbar^3V_5f_6'
-\hbar^2V_1f_6'
-4\hbar V_0'f_6
+4\hbar V_2'f_6
-4\hbar rV_3f_6=0.
\label{eq:det0_9}
\end{align}

\end{document}